\documentclass[prl,aps,twocolumn,groupaddress]{revtex4-1}
\usepackage{latexsym}
\usepackage{graphicx}
\usepackage{verbatim}
\usepackage{multirow}
\usepackage{amsmath}
\usepackage{mathrsfs}
\usepackage{float}
\usepackage[usenames, dvipsnames]{color}

\begin{document}

\title{Quasiparticle interference and resonant states in normal and superconducting line nodal semimetals}

\author{Chandan Setty}
\affiliation{Department of Physics, University of Illinois at Urbana-Champaign, Urbana, Illinois, USA}
\author{Philip W. Phillips}
\affiliation{Department of Physics, University of Illinois at Urbana-Champaign, Urbana, Illinois, USA}
\author{Awadhesh Narayan}
\affiliation{Department of Physics, University of Illinois at Urbana-Champaign, Urbana, Illinois, USA}
\affiliation{Materials Theory, ETH Zurich, Wolfgang-Pauli-Strasse 27, CH 8093 Zurich, Switzerland}

\begin{abstract}
We study impurity scattering in the normal and $d$-wave superconducting states of line nodal semimetals and show that, due to additional scattering phase space available for impurities on the surface, the quasiparticle interference pattern acquires an extended character instead of a discrete collection of delta function peaks. Moreover, using the $T$-matrix formalism, we demonstrate that the conventional behavior of a scalar impurity in a $d$-wave superconductor breaks down on the surface of a line nodal semimetal in the quasi flat band limit.
\end{abstract}

\maketitle

\textit{Introduction:} A recent member to the class of topological states~\cite{hasan2010colloquium,qi2011topological,bansil2016colloquium} of matter include line node semimetals, in which two bands are degenerate over an extended region have gapless excitations~\cite{burkov2011topological}. A number of materials have been proposed to exhibit such line node characteristics~\cite{xie2015new,PhysRevB.92.045108,PhysRevLett.115.036806,PhysRevLett.115.036807,PhysRevLett.115.026403,schoop2015dirac,neupane2016observation,wu2016dirac,hirayama2016topological,bian2016topological}, a list which is growing remarkably rapidly. These proposals, in turn, have inspired numerous theoretical studies of novel properties of this intriguing band structure~\cite{phillips2014tunable,rhim2015landau,koshino2016magnetic,ramamurthy2015quasi,yan2016tunable,narayan2016tunable,chan2016type,taguchi2016photovoltaic,wang2016topological,roy2016interacting}.

Inducing proximate superconductivity in topological states presents an intriguing playground for exotic forms of superconducting matter~\cite{zhang2011superconducting,sacepe2011gate,veldhorst2012josephson,yang2012proximity,wang2012coexistence}. Notably, high temperature proximity-induced superconductivity has been realized on canonical topological insulators bismuth selenide and bismuth telluride, using a $d$-wave cuprate superconductor~\cite{zareapour2012proximity}. Recent reports of tip-induced superconductivity in point node semimetals are an exciting new development in exploration of such phenomena~\cite{aggarwal2016unconventional,wang2016observation}.

At the same time, quasiparticle interference has proved to be an important tool in establishing and characterizing the fingerprints of topological matter. Surface states of topological insulators have been imaged and their spin-momentum locking has been revealed using scanning tunneling spectroscopy~\cite{roushan2009topological,zhang2009experimental,alpichshev2010stm,okada2011direct,alpichshev2012stm,honolka2012plane}. More recently, gapless topological phases of matter, Dirac and Weyl semimetals, have also been studied using scanning tunneling microscopy, where signatures of Fermi arcs have been found~\cite{jeon2014landau,inoue2016quasiparticle,zheng2016atomic,batabyal2016visualizing,sessi2017impurity}.

In this work, motivated by these advancements, we explore the quasiparticle interference in normal and superconducting line node semimetals focusing on both bulk and surface properties. We show that, unlike in conventional two dimensional metals where nodal superconductivity yields \textit{point} nodes, the surface of a line nodal semimetal gives rise to \textit{line} nodes. As a consequence, due to the additional impurity scattering phase space available within the area of the flat band, the quasiparticle interference pattern on the surface of a line nodal semimetal acquires an extended character in the Brillouin zone instead of a collection of discrete delta function peaks. Additionally, using the $T-$matrix formalism, we examine the resonant state energy dispersions of a single scalar impurity on the surface of a line nodal semimetal with $d-$wave pairing. Our calculations point to a momentum averaged Green function which contains a power law type contribution, in addition to the logarithmic term usually found for nodal superconducting quadratic bands. Such a contribution, unlike the case of two dimensional electrons with quadratic bands, admits two different under-damped solutions to the resonant state energies: the first is a broad, low intensity mode located closer to the continuum that disperses toward zero energy in the unitary limit; the second is a more intense, sharp, lower energy mode that disperses away from zero energy. We argue that first mode may be challenging to access experimentally while the second can be more readily observed.  Our results also signal a destruction of zero bias tunneling peaks (in the unitarity limit) on the surface of a line nodal semimetal with $d-$wave pairing and could, thereby, motivate future scanning tunneling experiments on line node semimetals. \\ \newline
\textit{Toy model for a line nodal semimetal:} To begin with, we briefly describe a slightly modified version of the tight binding toy model put forth in Ref.~\cite{Schnyder2016} and study some of its bulk and surface properties. Equipped with a basic understanding of these properties, we go on to study the impurity induced quasiparticle interference patterns in both the normal and superconducting states of the line node semimetal. We take our tight binding Hamiltonian on a square lattice to be of the form (we use the same notation as in Ref.~\cite{Schnyder2016} to make the comparison explicit)

\begin{figure}[h!]
\includegraphics[width=1.7in,height=1.9in]{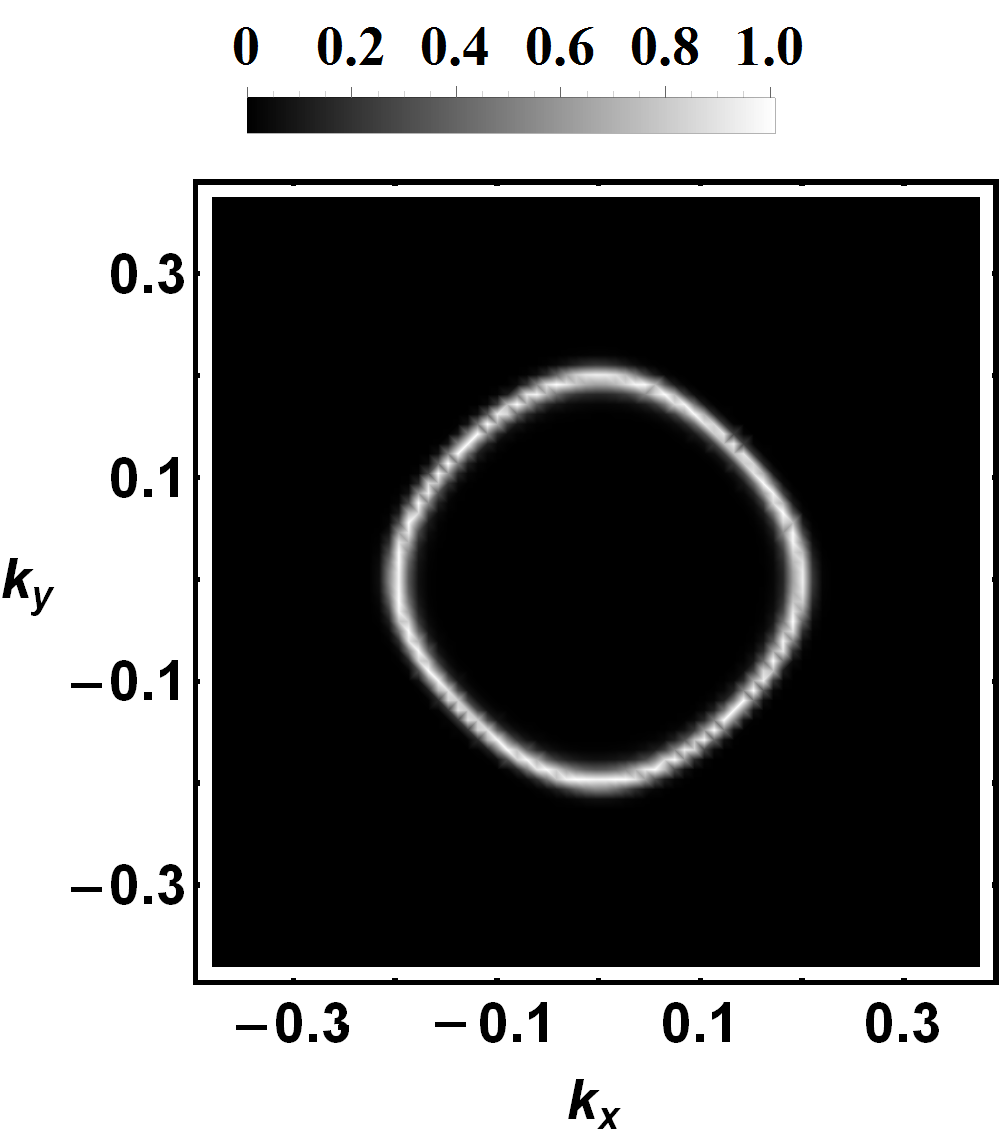}\hfill%
\includegraphics[width=1.7in,height=1.9in]{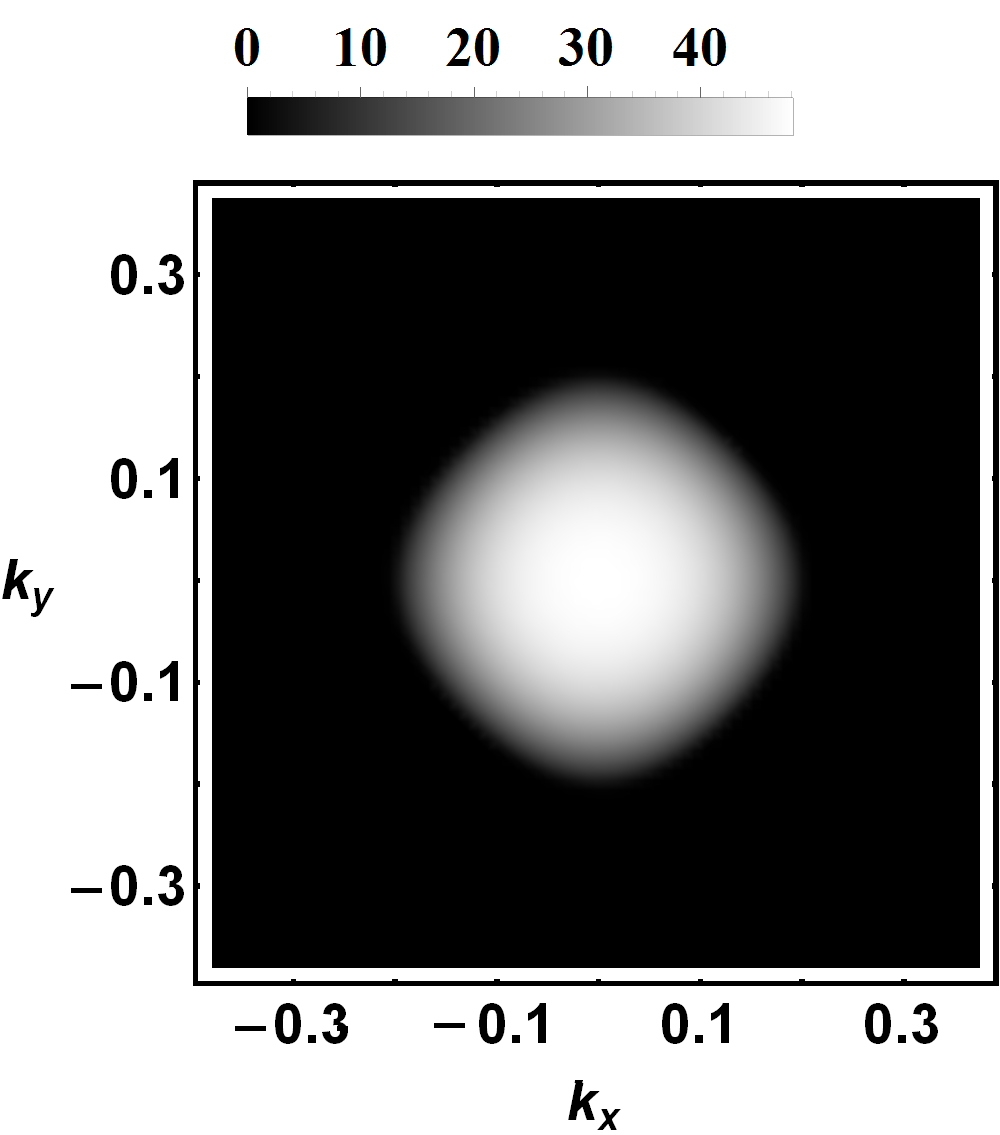}\hfill %
\includegraphics[width=1.7in,height=1.9in]{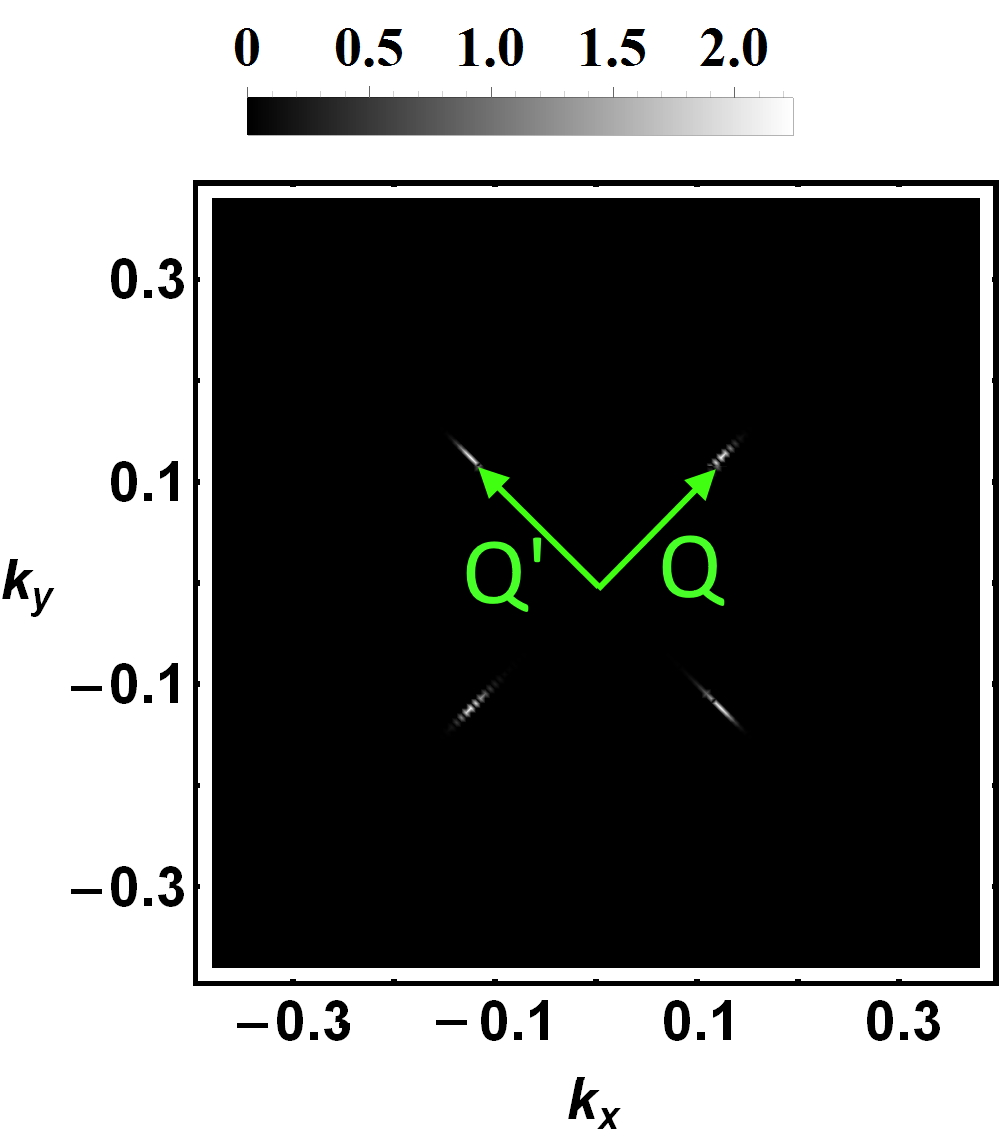}\hfill%
\includegraphics[width=1.7in,height=1.9in]{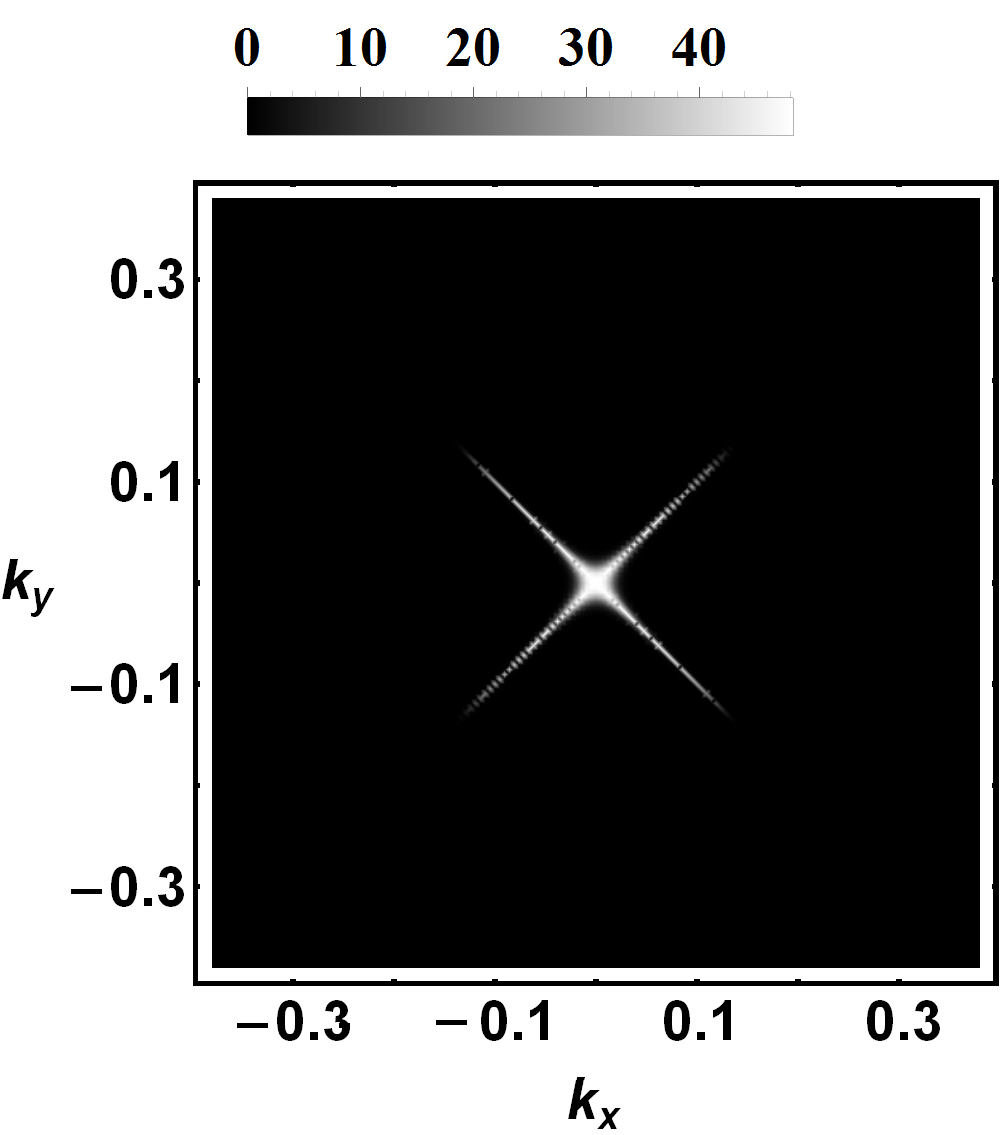}
\caption{Zero energy local density of states for the line node semimetal. Top row: (Left) Bulk and (right) Surface without superconductivity. Bottom row: (Left) Bulk and (right) Surface with $d-$wave superconductivity.}
\label{SchnyderModelZeroOmega}
\end{figure}

\begin{eqnarray}\nonumber
\hat{H}_0(\vec k) &=& \left[\frac{g(\vec k_{\parallel})\nu_{\parallel}'}{a^2} \tau_z + \left(\frac{\bar{g}\nu_{0}'}{a^2} + V_0 \right)\tau_0\right] \sigma_0\\ \label{Hamiltonian}
&&+\frac{\nu_z}{c} \sin(ck_z) \tau_y \sigma_0 + \hat{H}_z\\
\hat{H}_z&=& \left(1-\cos (c k_z)\right)\left(Z_{\tau}\tau_z \sigma_0 + Z_0 \tau_0\sigma_0\right)
\end{eqnarray}

\noindent where $\tau_i, \sigma_i$ are the Pauli matrices in the orbital and spin basis respectively and $a,c$ are the in-plane and out-of-plane lattice constants. The function $g(\vec k_{\parallel})$ is defined as $g(\vec k_{\parallel}) = 1 + \cos(ak_0) - \cos(ak_x) - \cos(ak_y)$. We set the parameters to the following values $\left(Z_{\tau},Z_0, a,c, k_0, V_0\right)$=(0.287 eV, 0.0 eV, 8.26 \AA, 6.84 \AA, 0.206 \AA$^{-1}$, 0.043 eV) and define $\nu_{\parallel}' = \frac{2 \nu_{\parallel} a k_0}{\sin(ak_0)}$, $\nu_{0}' = \frac{2 \nu_{0} a k_0}{\sin(ak_0)}$, $\bar{g} = 1+ \cos(ak_0)$ with $(\nu_0,\nu_{\parallel},\nu_z)$=(-0.993 eV\AA$^2$, 4.34 eV\AA$^2$, 2.5 eV\AA). To explore the flat band surface states, we use open boundary conditions along one of the directions, namely the $z$ axis.


In Fig.\ref{SchnyderModelZeroOmega}, we plot the local density of states (LDOS) in the bulk and surface of the model described in Eq.~\ref{Hamiltonian} at zero energy. The top row shows the LDOS in the bulk (left) and surface (right) in the normal state. In the bulk, there is a continuous contour of Dirac nodes at zero energy which acquires a toroidal structure at non-zero frequencies (see Supplemental Material~\cite{supplement}).  However, on the surface, a nearly flat band is found which ``fills'' the bulk contour, the so-called ``drumhead states''. In our discussions, we will be most interested in the flat band limit of the model where the surface band dispersion is the smallest energy scale in the problem. At non-zero energy, the flat-band behavior on the surface is absent and there is little qualitative distinction between the surface and the bulk~\cite{supplement}. It is also worthwhile to note that it is possible to augment the Hamiltonian in Eq.~\ref{Hamiltonian} to include terms which smoothly interpolate between a line nodal semimetal and a Weyl semimetal (see Supplemental Material~\cite{supplement}).

\begin{figure*}[t!]
\includegraphics[width=1.7in,height=1.65in]{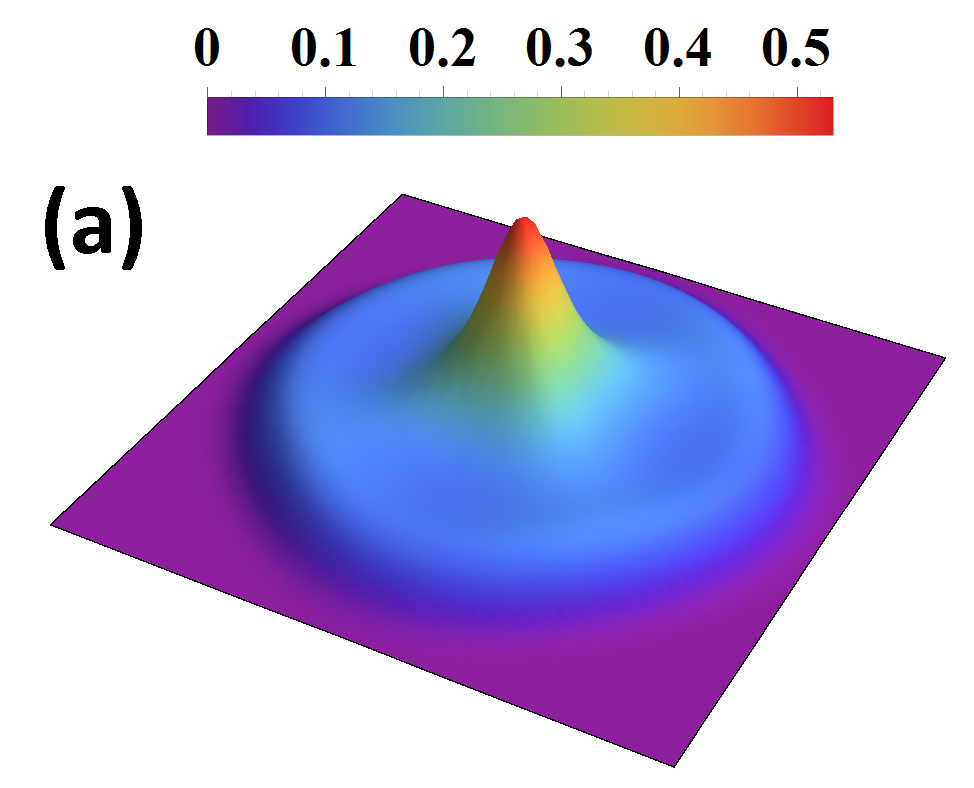}\hfill%
\includegraphics[width=1.7in,height=1.65in]{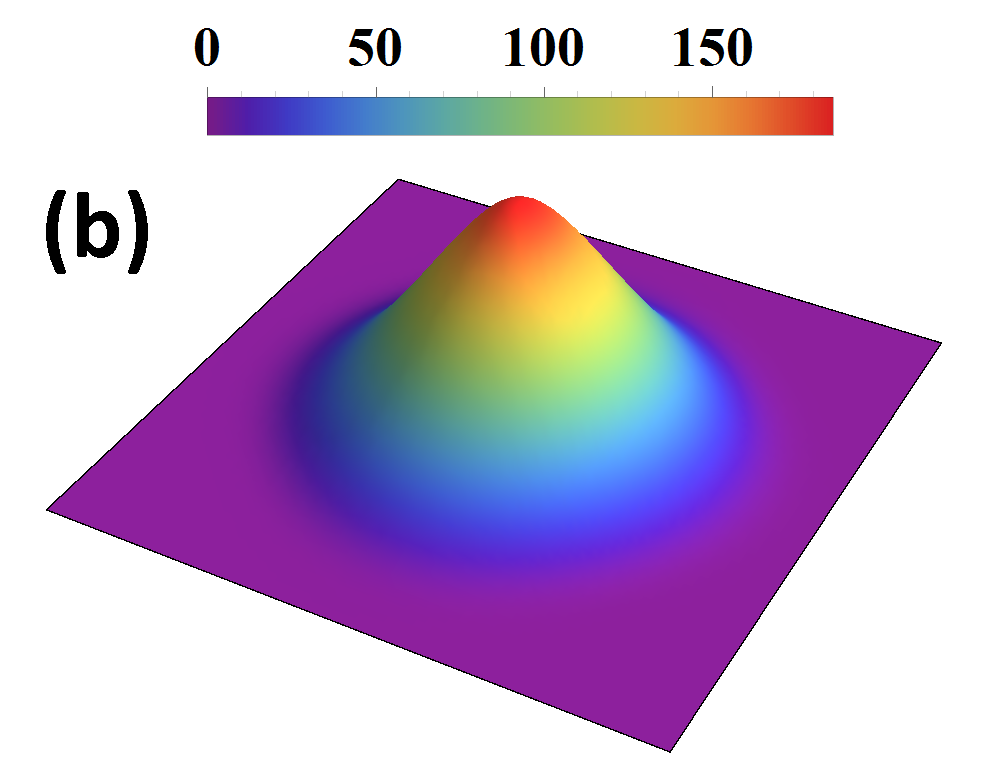}\hfill %
\includegraphics[width=1.68in,height=1.7in]{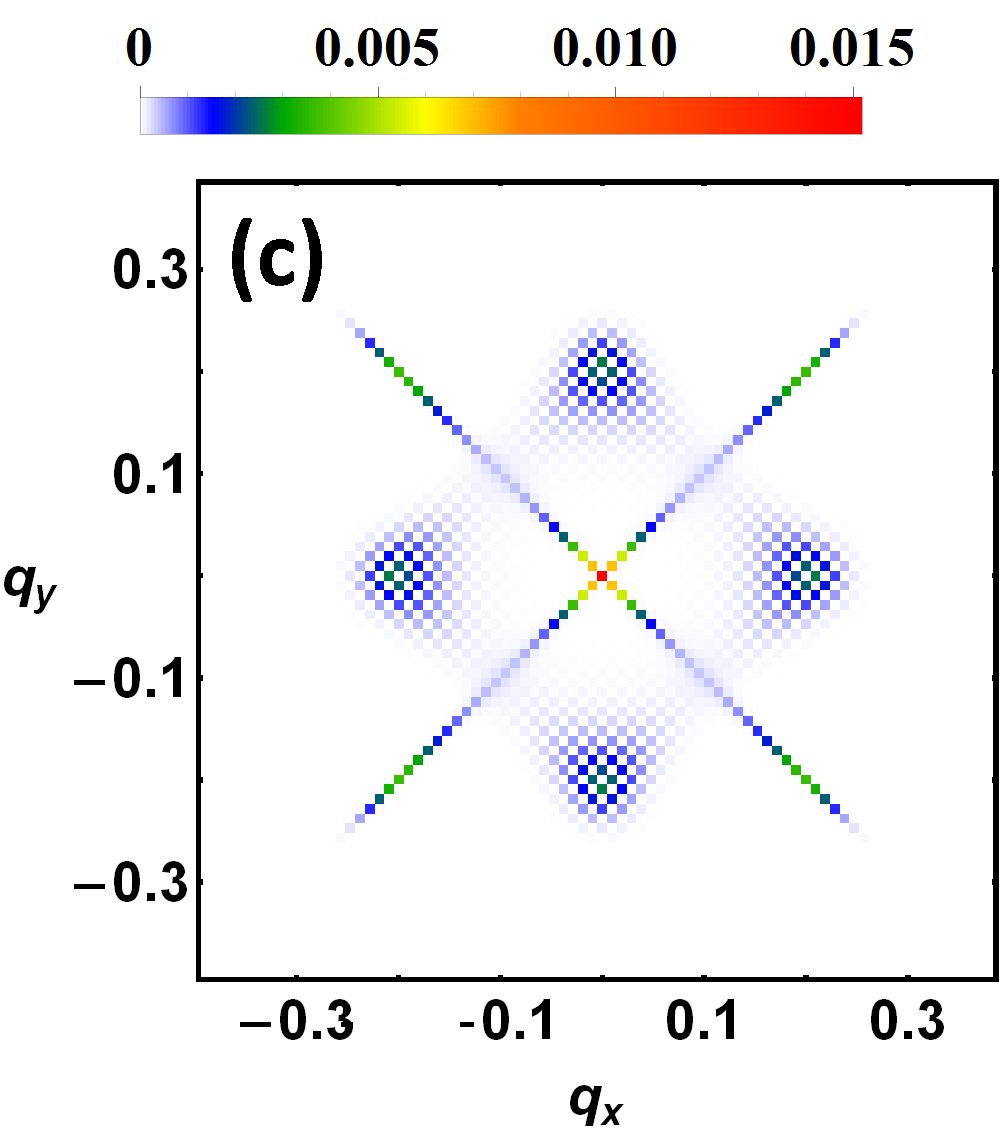}\hfill%
\includegraphics[width=1.68in,height=1.7in]{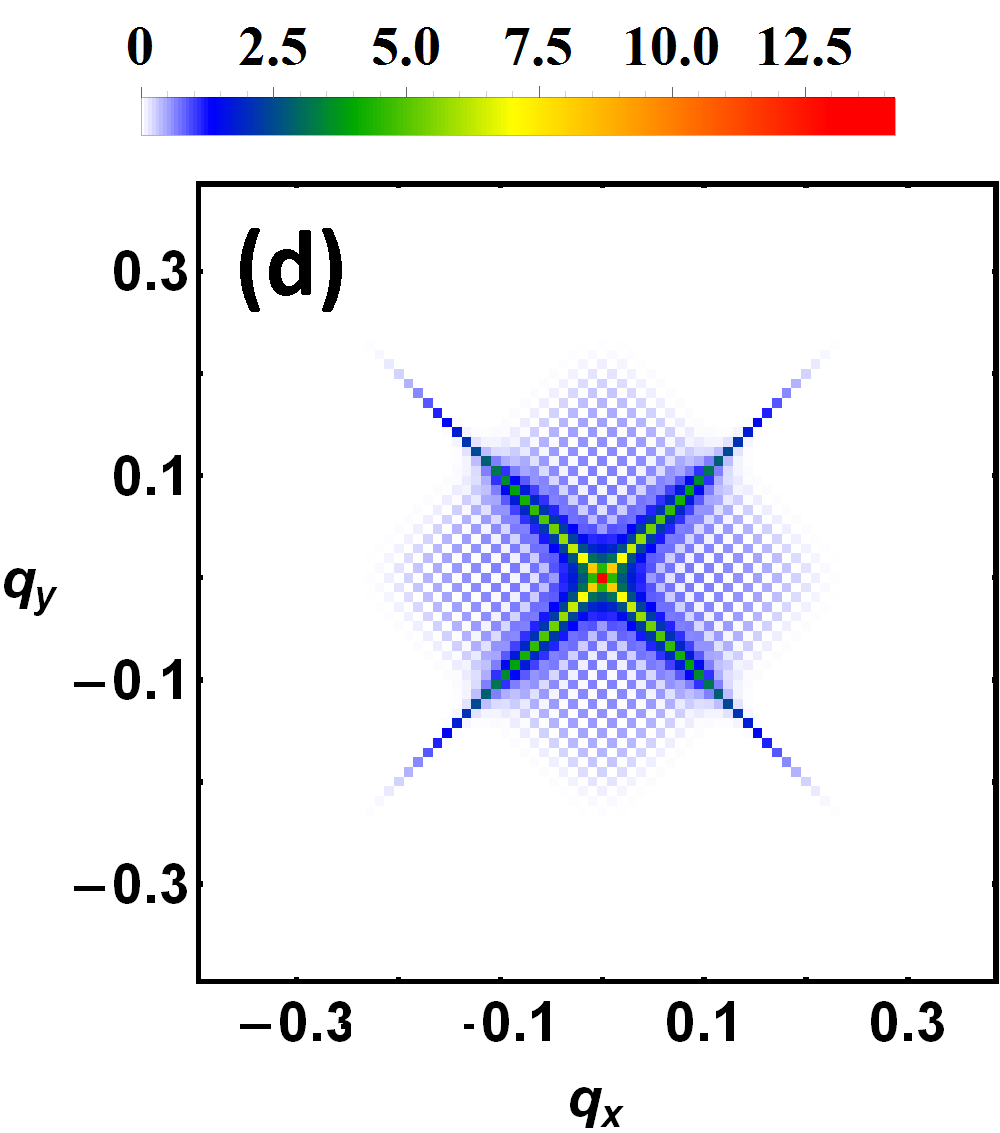}\hfill%
\includegraphics[width=1.7in,height=1.65in]{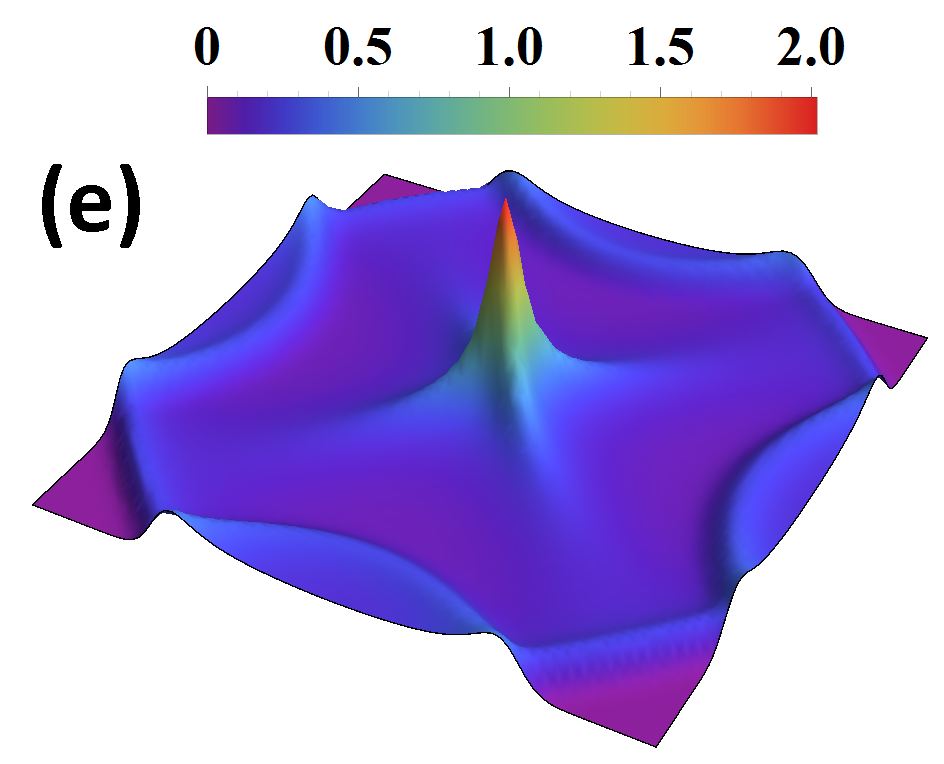}\hfill%
\includegraphics[width=1.7in,height=1.65in]{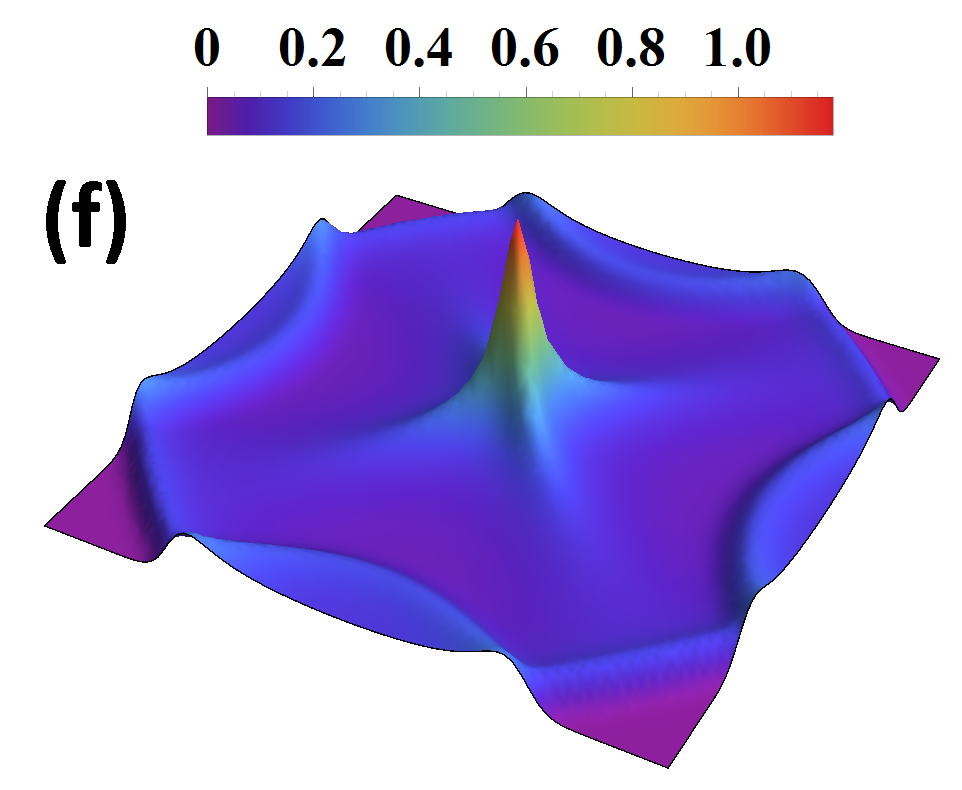}\hfill %
\includegraphics[width=1.68in,height=1.7in]{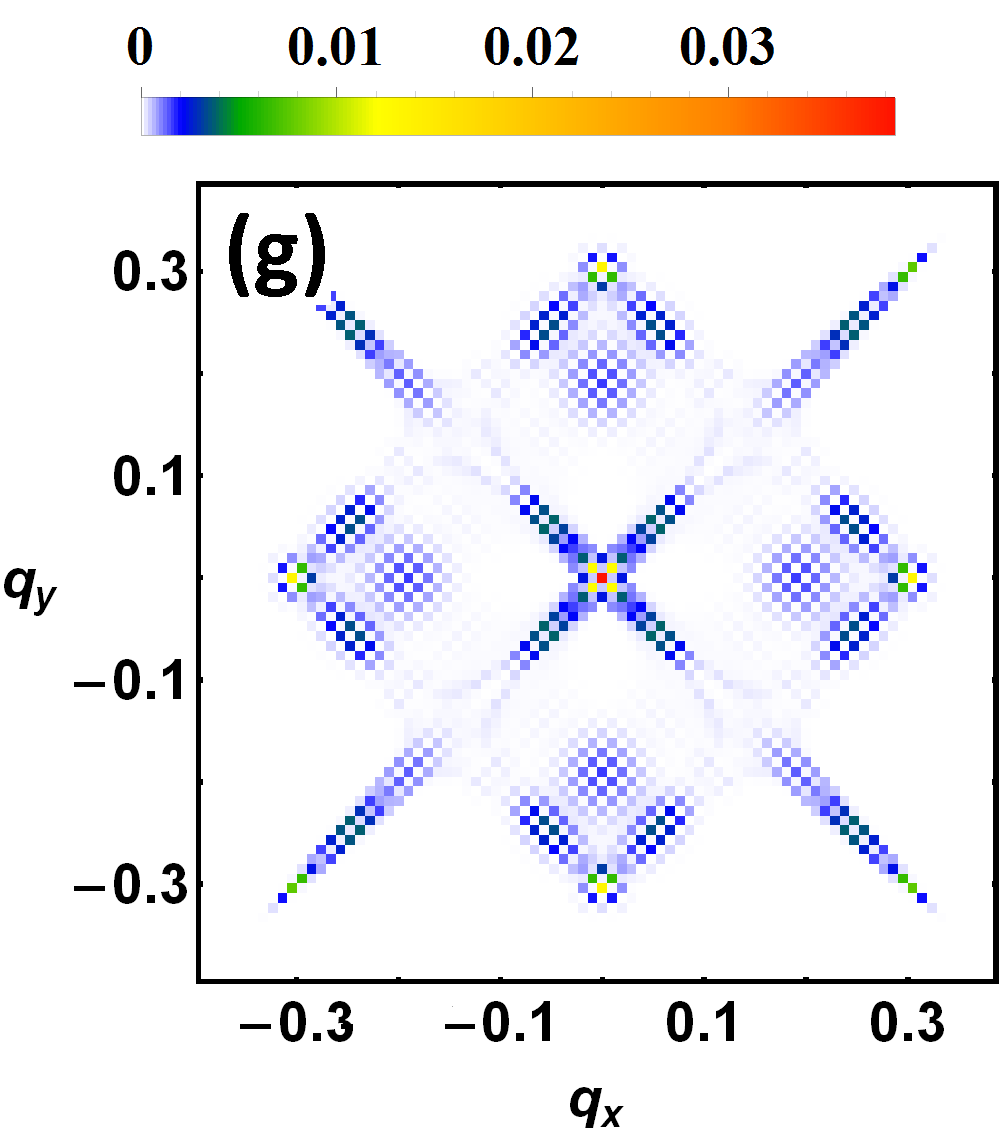}\hfill%
\includegraphics[width=1.68in,height=1.7in]{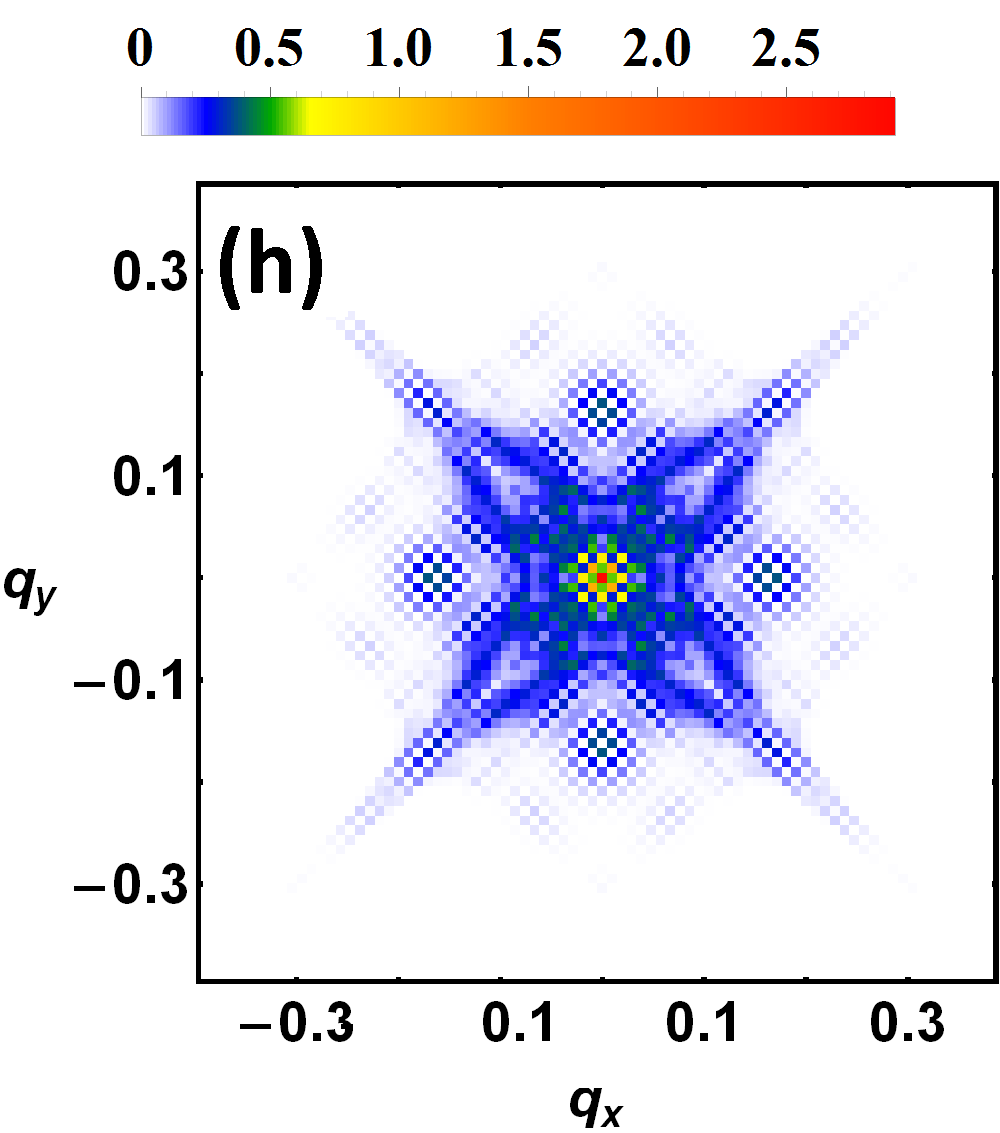}
\caption{Joint density of states at $\omega=0$ (panels a,b,c,d) and $\omega=0.1$eV (panels e,f,g,h). Panels on the left  half (a,b,e,f) are in the normal state and the right half (c,d,g,h) are in the $d-$wave superconducting state. Panels a,e,c and g correspond to the bulk JDOS and b,f,d,h correspond to the surface JDOS. The $q_x$ and $q_y$ axes on the left half of the figure have the same range as panels on the right half.}
\label{JDOSZeroOmega}
\end{figure*}

We now include proximity-induced superconductivity in our setup (see~\cite{supplement} for the details). For the rest of the paper, we will model the superconductor in the even frequency, orbital and spin singlet pairing channel. Noting that fully gapped $s-$wave superconductors are robust and featureless to scalar impurities due to Anderson's theorem, interesting impurity effects start to appear with nodal $d-$wave pairing, which will be the main focus of this work. The effect of a $d-$wave form of the gap on the bulk and surface LDOS at zero frequency is shown in Fig.~\ref{SchnyderModelZeroOmega} (bottom row). The intensity in the bulk (Fig.~\ref{SchnyderModelZeroOmega} bottom left) is now reduced to four nodal spots corresponding to the zeros of the $d-$wave gap function. These nodal points are marked by arrows denoted by $\vec Q$ and $\vec Q'$. However, on the two dimensional surface (Fig.~\ref{SchnyderModelZeroOmega} bottom right) a $d-$wave gap gives rise to line nodes instead of point nodes$-$ a novel feature of the ``drumhead'' surface state that does not occur in usual two dimensional superconductors. This would lead to an anomalous scaling of measurable quantities, like the specific heat, leading to a striking difference which could be readily tested in future experiments.

At non-zero energies, the four nodal points that existed in the bulk become slightly extended in momentum space (see Supplemental Material~\cite{supplement}) along the diagonals of the Brillouin zone due to the toroidal Fermi surface. On the surface, however, when the induced superconducting gap ($\Delta_{\mu\nu}$) is larger than the chosen energy ($\omega = 0.1$ eV and $\Delta_{\mu\nu}>0.1$ eV), then all the surface bands become gapped \textit{except} those states along the Brillouin zone diagonal. These states then converge down to the Fermi level to form line nodes at zero energy, while at non zero energy (less than the maximum value of superconducting gap) they form a ``petal'' like structure.  \\ \newline 
\textit{Impurity scattering:} With the analysis of LDOS in the normal and superconducting phases at hand, we are now in a position to examine the effect of impurity scattering on line node semimetals. In the presence of impurities, the electrons in states with high density at the same energy can scatter between these states. This gives rise to interference patterns which can be measured using scanning tunneling methods. Joint density of states (JDOS) has proved to be a useful quantity to compare to experimentally obtained quasiparticle interference patterns and to analyze the possible scattering processes~\cite{roushan2009topological}. It can be obtained in a straightforward manner by $\mathrm{JDOS}(\vec q,z)=\int \mathrm{DOS}(\vec k,z)\mathrm{DOS}(\vec k+\vec q, z)d^2\vec k$. The simplicity of the computation then allows a detailed analysis of the obtained interference pattern. 

Fig.~\ref{JDOSZeroOmega} shows the JDOS at $\omega=0$ (Fig.~\ref{JDOSZeroOmega} panels a-d ) and $\omega=0.1$eV (Fig.~\ref{JDOSZeroOmega} panels e-h ) in the normal (panels a,b,e,f) and superconducting states (panels c,d,g,h). For $\omega=0$ in the normal state, both the bulk (panel a) and surface (panel b) JDOS show dominant peaks at the Brillouin zone center corresponding to impurity scattering with zero momentum. The 'radius' of the region with non-zero JDOS intensity for both the cases is about twice that of the vectors $\vec Q$ and $\vec Q'$, as is expected from scattering between these states. However, there are some important features that distinguish the surface and the bulk JDOS even without induced superconductivity. First, the intensity of the JDOS is much larger on the surface than in the bulk (at zero energy) due to the surface flat band. Second, the JDOS profile in the bulk (Fig.~\ref{JDOSZeroOmega}(a) ) is quasi-flat away from zero momentum transfer and peaks steeply at zero momentum. On the other hand, the surface JDOS (Fig.~\ref{JDOSZeroOmega}(b) ) has a thick cone like feature. This difference is due to the additional impurity scattering contributions originating from all the momenta within the boundary of the surface flat band which is absent in the bulk. 

 In the presence of induced $d$-wave superconductivity (Fig~\ref{JDOSZeroOmega} panels c,d,g,h) at zero energy (panels c and d), the bulk (panel (c)) JDOS profile essentially peaks at nine points in the Brillouin zone. These points correspond to $\vec q = 0, \pm 2 \vec Q, \pm 2 \vec Q', \pm (\vec Q + \vec Q'), \pm (\vec Q - \vec Q') $ which represent the nine different ways to connect the four nodal spots with themselves and with the rest of the others  (see Fig.~\ref{SchnyderModelZeroOmega} bottom, left). The surface JDOS (panel (d)) in the presence of induced superconductivity has additional intensity within the square bounded by the momentum vectors $\pm (\vec Q + \vec Q'), \pm (\vec Q - \vec Q') $. This is entirely a consequence of the fact that $d-$wave superconductivity yields line nodes on the surface of a line nodal semimetal instead of point nodes (as in the bulk). In such a scenario, all the momentum vectors that lie within the square, correspond to vectors that connect different points on the X shaped line node (in the DOS appearing in Fig.~\ref{SchnyderModelZeroOmega} bottom, right) with each other. This is strikingly different from the situation in $d$-wave superconductivity in materials lacking the ``drumhead'' states. This could prove to be an experimentally verifiable signature of the surface states of line node semimetals. As discussed before, at non-zero energies, there is little difference between the bulk (panel e) and the surface (panel f) in the absence of superconductivity; in fact, the surface has a smaller JDOS intensity than the bulk due to the absence of surface states away from the Fermi level. In the presence of $d-$wave superconductivity, however, the bulk (panel g) and surface (panel h) JDOS profiles start to acquire broadened characteristics in accordance with the LDOS. In such a case, the surface still has a greater intensity than the bulk because surface states with momenta along the diagonals disperse all the way down to zero energy.  \\ 
 
\begin{figure*}[t!]
\includegraphics[width=2.35in,height=1.6in]{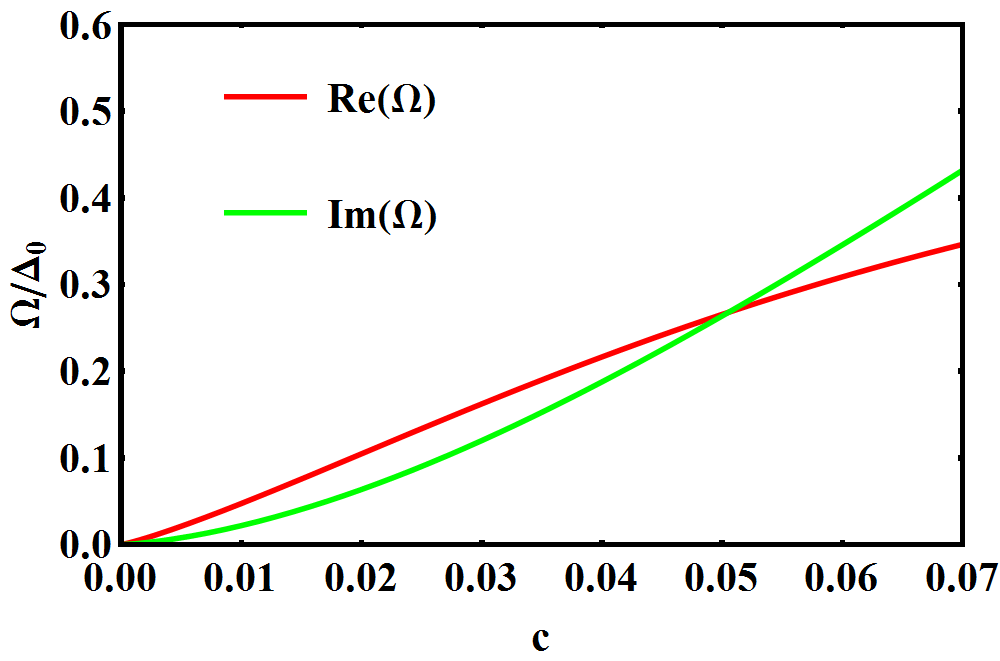}\hfill%
\includegraphics[width=2.35in,height=1.6in]{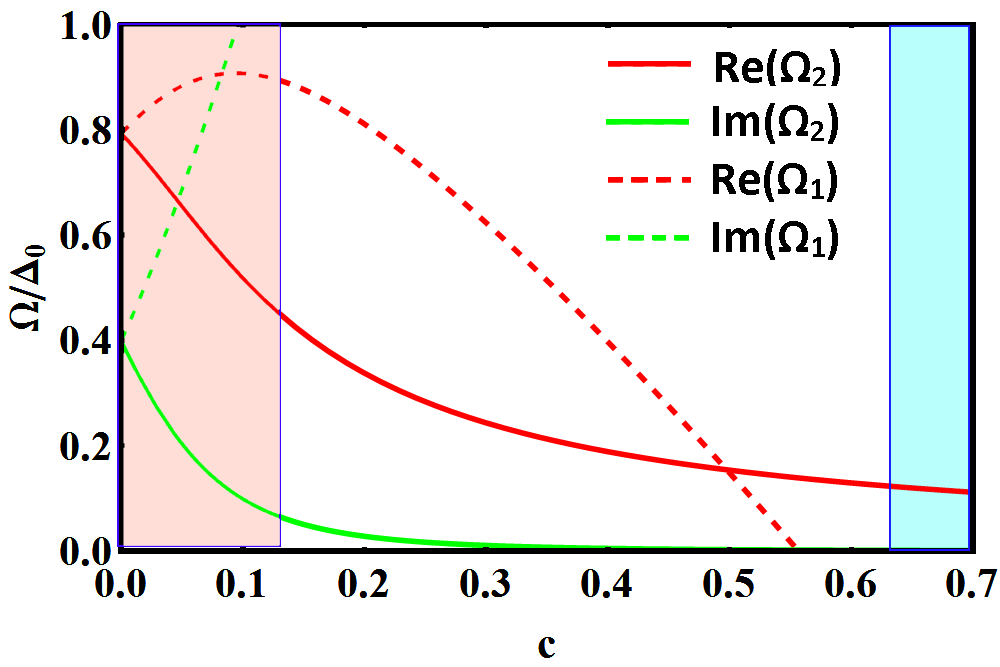}\hfill%
\includegraphics[width=2.35in,height=1.6in]{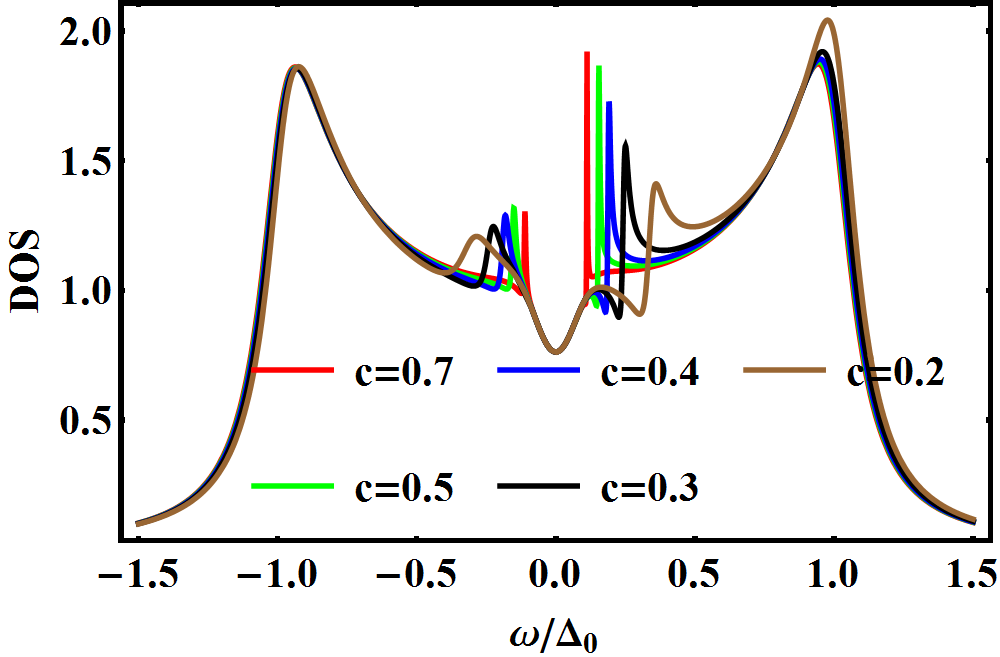}%
\caption{Comparison of the real and imaginary parts of the resonant state energies obtained by solving $\mathscr G_0(\Omega) = \pm c$. (Left) Without the $\frac{1}{\xi^2}$ term in Eq.~\ref{G-QFB}. This is similar to the case of a $d$-wave superconductor with a quadratic dispersion. (Center) The case corresponding to the quasi-flat band where there are two solutions $\Omega_{1,2}$ admissible. There is a regime for small $c$ (left shaded) where both the resonances$-$though well defined$-$are broad and have low spectral intensity; hence, they are challenging to observe experimentally. In the opposite limit (right shaded), the $\Omega_2$ solution no longer holds due to weakening of the flat band approximation. (Right) Corresponding DOS vs energy plots. Note that for these values of $c$, $\Omega_1$ is damped.}
\label{QFB}
\end{figure*}
 
\textit{Impurity resonant states and $T$-Matrix approximation:} Next, we analyze resonant states that may arise around the impurities in line node semimetals.To clarify the notation, we  briefly outline the $T-$matrix approximation (for further details refer to~\cite{Zhu2006}). The total electron Green function is written as
\begin{equation}
\hat G(\vec k, \vec k', \omega) =\hat G_0(\vec k, \omega) \delta_{\vec k, \vec k'} + \hat G_0(\vec k,\omega) \hat T(\vec k,\vec k', \omega) \hat G_0(\vec k', \omega),
\end{equation}
where $G(\vec k, \vec k', \omega)$ and $G_0(\vec k, \omega)$  are the total interacting and non-interacting Green functions, and $T(\vec k, \vec k', \omega)$ is the $T-$matrix  which contains the physics originating from impurity scattering. For the purposes of this article, we confine ourselves to scalar potential scatterers; this renders the $T-$matrix momentum independent and can be written as $\hat T(\omega)=\left[\hat{\sigma}_0 - \hat V \hat g_0(\omega)\right]^{-1}\hat V$. Here, we have defined $\hat g_0(\omega) = \frac{1}{2\pi N_0}\sum_{\vec k} \hat G_0(\vec k, \omega)$, with $N_0$ being the density of states at the Fermi level, and the scattering matrix $\hat V$ given by $\frac{1}{c}\hat{\tau}_3$. We have also used the parameter $c = \cot(N_0 U_0)$ as a measure of the strength of an isotropic scatterer, following Ref.~\cite{Einzel1988}, where $U_0$ is the strength of the impurity scatterer. Therefore, the unitarity limit (large scattering strength, $N_0 U_0\rightarrow \frac{\pi}{2}$) corresponds to the case when $c\rightarrow 0$.\\ \newline
Before we move on to the superconducting state of a line-nodal semimetal, we briefly recall known results regarding resonant state dispersions of a scalar impurity in $d-$wave superconductors from the works of Balatsky and Hirschfeld~\cite{Einzel1988,Rosengren1995,Zhu2006}. We begin by writing out the non-interacting Greens' function given as $\hat G_0(\vec k,\omega) = \left( \omega \hat{\sigma}_0 - \hat H_{sc}(\vec k)\right)^{-1}$, where $\hat H_{sc}(\vec k) = \epsilon(\vec k) \hat{\sigma}_3 + \Delta(\vec k)\hat{\sigma_1}$, $\epsilon(\vec k) = \alpha k^2 - \mu$ ($\alpha$ is a constant and $\mu$ is the chemical potential), and $\Delta(\vec k) =  \Delta_0 \cos 2\phi_{\vec k}$. In general, the matrix $\hat g_0(\omega)$, can be written as $\hat g_0(\omega) = \sum_{i}\mathscr G_i(\omega) \hat{\sigma}_i $ where we have $\mathscr G_i(\omega)\equiv \frac{1}{2 \pi N_0}\sum_{\vec k} \mathscr G_i(\omega,\vec k)$, $\mathscr{G}_0(\omega, \vec k) = \frac{-\omega}{D_{\vec k}}$, $\mathscr{G}_1(\omega, \vec k) = \frac{-\Delta(\vec k)}{D_{\vec k}}$, $\mathscr{G}_2(\omega,\vec k) =0$, $\mathscr{G}_3(\omega, \vec k) = \frac{-\epsilon(\vec k)}{D_{\vec k}}$ and $D_{\vec k} = \Delta(\vec k)^2 + \epsilon(\vec k)^2 - \omega^2$. Given the form of the scattering matrix, $\hat V = \frac{1}{c}\hat{\sigma}_3$, the condition for the existence of resonant states is that the determinant of $\left[ \hat{\sigma}_0 - \hat V \hat g_0(\omega) \right]$ must vanish. This translates to

\begin{equation}
\mathscr G_1(\omega)^2 - \mathscr G_0(\omega)^2 + (c - \mathscr G_3(\omega))^2 =0.
\label{Condition}
\end{equation}

Our task now is to evaluate these functions for the case of a $d-$wave superconductor with a quadratic dispersion in two dimensions. The quantity $\mathscr G_1(\omega)$ is zero since the gap function changes sign across the Brillouin zone and the $\phi$ integral vanishes. Similiary $\mathscr{G}_3(\omega)$ is zero if we assume particle-hole symmetric bands in two dimensions. Keeping this in mind, we evaluate $\mathscr G_0(\omega)$ for quadratic bands and, in the limit $\omega\ll \Delta_0$, it can be shown that \cite{Einzel1988}
\begin{equation}
\mathscr G_0(\omega) \simeq \frac{-\omega}{\pi \Delta_0}\left[ \log\left(\frac{4 \Delta_0}{\omega}\right) - i \frac{\pi}{2}\right].
\label{2DQuadraticGreen}
\end{equation}

The condition for the existence of a resonant state (appearing in Eq.~\ref{Condition}) with frequency $\Omega$ (whose real and imaginary parts are denoted by $\mathrm{Re}(\Omega)$ and $\mathrm{Im}(\Omega)$) simply reduces to $\mathscr G_0(\Omega) = \pm c$. The only under-damped solution to this equation as a function of $c$ has two important features to which one needs to pay attention (Fig~\ref{QFB} left panel): (i) both the real and imaginary parts of $\Omega$ go to zero in the unitarity limit ($c\rightarrow0$).  This implies that the in- gap resonant state gets sharper and softer (yields a sharp zero bias peak in the unitarity limit) as a function of the impurity scattering strength and (ii) for a finite range of $c$, the real part of $\Omega$ is slightly larger than the imaginary part of $\Omega$. This is the regime where the resonant state is reasonably well defined, and above this value of $c$, the state is heavily damped. We wish to compare this result to the dispersive properties of an impurity on the surface of a line nodal semimetal with a $d-$wave pairing in the quasi-flat band limit. To do so, we choose the normal state density of states profile as a Lorenztian of the form $\rho(\epsilon) = \frac{\gamma/\pi}{\gamma^2 + \epsilon^2}$ with a width $\gamma$ that peaks at the Fermi level. The energy scale $\gamma$ can be chosen to be the smallest among all other energy scales in the problem (bandwidth $W$, pairing amplitude $\Delta_0$ and frequency $\omega$). Similar to the previous case of a quadratic dispersion, we have $\mathscr G_1(\omega)$ and $\mathscr G_3(\omega)$ to be zero. To calculate $\mathscr G_0(\omega)$ for the surface of a line nodal semimetal, we substitute for the Lorentzian density of states profile into the momentum integral. In the limit of $W\gg\Delta_0\gg\omega\gg\gamma$, we obtain (see Supplemental material for details~\cite{supplement})

\begin{equation}
\mathscr G_0(\omega)_{LNS} \simeq -\frac{2\gamma^2\xi}{\Delta_0^2} \left[\frac{1}{\xi^2} +\frac{1}{2} \log\left(\frac{\xi}{4}\right) + i\frac{\pi}{4}\right], 
\label{G-QFB}
\end{equation}

\noindent where we have defined $\xi\equiv\frac{\omega}{\Delta_0}$. This form of $\mathscr G_0(\omega)_{LNS}$ bears some similarities to the ones we derived in Eq.~\ref{2DQuadraticGreen}; however, the crucial difference in Eq.~\ref{G-QFB} is the appearance of an additional term $\frac{1}{\xi^2}$ due to the presence of the quasi-flat band. 
This power law term has important consequences to the resonant state energies (see Fig.~\ref{QFB}). Unlike the two dimensional electron case with quadratic bands, the condition $\mathscr{G}_0(\omega) = \pm c$ admits two under-damped solutions ($\Omega_{1,2}$), one for each sign. The real and imaginary parts of these solutions are shown in the center panel of Fig.~\ref{QFB}. While $\Omega_1$ is weakly undamped only for small $c$, $\Omega_2$ remains sharp for all values of $c$. Moreover, the real parts of $\Omega_1$ and $\Omega_2$ disperse in opposite directions in the unitary regime. Note, however, that the dispersion of the real part of $\Omega_2$  cannot go on to zero energy in the weak scattering (or large $c$) limit. It is reasonable to expect this as there should be no in-gap resonant states when the scattering strength goes to zero. Our result is consistent with this expectation since for large values of $c$ (shaded region on the right in Fig.~\ref{QFB}, center panel), $\Omega_2$ becomes comparable to $\gamma$, and the quasi-flat band approximation weakens and eventually breaks down. On the other hand, in the unitarity limit $c\lesssim0.1$ (shaded region on the left in Fig.~\ref{QFB}, center panel), the real parts of both $\Omega_1$ and $\Omega_2$ approach a relatively large fraction ($\frac{\Omega}{\Delta_0}\sim 0.8$; compare this to the quadratic band case in Fig.~\ref{QFB} left most panel, where it goes to zero energy) of the maximum gap value. This proximity to the continuum, coupled with the fact that the peak intensities go to zero for large impurity scattering, makes it experimentally challenging to observe this mode. Therefore, there is an optimal window of the scattering strengths where the resonance occurs predominantly due to quasi-flat band effects and, at the same time, is experimentally observable (see Fig.~\ref{QFB}, right panel). Finally, there is expected to be little spatial variation of the peak intensity on different sites close to/ at the impurity~\cite{andersen2006andreev, kreisel2015interpretation} due to lack of spatial dynamics in a quasi-flat band system.  \newline 
\textit{Summary:} To conclude, we studied the effect of nodal $d-$wave pairing in the bulk and on the surface of a line nodal semimetal, and determined the role of impurities through the joint density of states, which could be measured via quasiparticle interference experiments. We observed that, unlike conventional two-dimensional metals where nodal superconductivity yields \textit{point} nodes, the surface of a line nodal semimetal gives rise to \textit{line} nodes. As a consequence, due to the additional impurity scattering phase space available within the area of the flat band, the JDOS pattern on the surface of a line nodal semimetal acquires an extended character in the Brillouin zone instead of a collection of discrete delta function peaks. Using the $T-$matrix formalism, we also examined resonant state energy dispersions of a single scalar impurity on the surface of a line nodal semimetal with $d-$wave pairing. Our results demonstrated that the momentum averaged Green function contains a power law type contribution in addition to the logarithmic term usually found for nodal superconducting quadratic bands.  Such a contribution admits two different under-damped solutions to the resonant state energies, unlike the case of two dimensional electrons with quadratic bands where there is only one under-damped solution. The first solution is a broad, low intensity mode located closer to the continuum that disperses toward zero energy in the unitary limit; the second is a more intense, sharp, lower energy mode that disperses away from zero energy. We argued that first mode may be challenging to access experimentally while the second can be more readily observed. Our results also signal a destruction of zero bias tunneling peaks (in the unitarity limit) on the surface of a line nodal semimetal with $d-$wave pairing.  Looking forward, it could be interesting to explore impurity effects in Josephson junctions on line node semimetal surfaces, analogous to investigations on helical metals \cite{ghaemi2016effect}. We are hopeful that our findings would motivate scanning tunneling spectroscopic experiments on line node semimetals.\newline
\textit{Acknowledgments:} We acknowledge support from Center for Emergent
Superconductivity, a DOE Energy Frontier Research Center, Grant No. DE-AC0298CH1088. A.N. acknowledges additional support from ETH Zurich.

\bibliographystyle{apsrev4-1}
\bibliography{Impurity}

\newpage

\title{Quasiparticle interference and resonant states in normal and superconducting line nodal semimetals}

\author{Chandan Setty}
\affiliation{Department of Physics, University of Illinois at Urbana-Champaign, Urbana, Illinois, USA}
\author{Philip W. Phillips}
\affiliation{Department of Physics, University of Illinois at Urbana-Champaign, Urbana, Illinois, USA}
\author{Awadhesh Narayan}
\affiliation{Department of Physics, University of Illinois at Urbana-Champaign, Urbana, Illinois, USA}
\affiliation{Materials Theory, ETH Zurich, Wolfgang-Pauli-Strasse 27, CH 8093 Zurich, Switzerland}
\onecolumngrid 
\section{SUPPLEMENTAL MATERIAL}

\textbf{\textit{Bogoliubov-de Gennes equations:}}
For the inclusion of proximity induced superconductivity and the joint density of states in our setup appearing in the main text, and for the purposes of fixing our notation, we provide a basic introduction to the Bogoliubov-de Gennes (BdG) equations. For the superconducting state with periodic boundary conditions along all the three directions, we use the BdG Hamiltonian in momentum space given by~\cite{de1989superconductivity}

\begin{eqnarray}
H_{BdG} &=& \sum_{\vec k\mu \nu}  
\begin{pmatrix}
 c^{\dagger}_{\vec k \uparrow\mu} & c_{-\vec k \downarrow \mu}
 \end{pmatrix}
 \hat{H}_{BdG}(\vec k) 
\begin{pmatrix}
 c_{\vec k \uparrow\nu} \\ c^{\dagger}_{-\vec k \downarrow\nu}
 \end{pmatrix}\\
 \hat{H}_{BdG}(\vec k)  &=& 
 \begin{pmatrix}
 H_0(\vec k)_{\mu \nu} & \Delta_{\mu \nu}  \\
\Delta^{\dagger}_{\mu \nu} &  -H_0(-\vec k)_{\mu \nu}
 \end{pmatrix},
\end{eqnarray}

\noindent where $c_{\vec k \sigma \nu}$ and $c_{\vec k \sigma \nu}^{\dagger}$ are the annihilation and creation operators for electrons in orbital $\nu$, momentum $\vec k$ and spin $\sigma$ and $\Delta_{\mu\nu}$ is the induced superconducting gap. In the presence of a surface, when we have open boundary conditions along the $z-$direction we can only Fourier transform along the $k_x$ and $k_y$ directions (together denoted as $\vec k_{\parallel}$). In such a scenario one can generically write the BdG eigenvalue equation as
\begin{equation}
\hat{H}_{BdG}(\vec k_{\parallel}, z) |\psi_n(\vec k_{\parallel}, z)\rangle=  E_n(\vec k_{\parallel}, z) |\psi_n(\vec k_{\parallel}, z)\rangle
\end{equation}

\noindent where $\hat{H}_{BdG}$ is the BdG Hamiltonian Fourier transformed only along the $\vec x_{\parallel}$ direction, $|\psi_n(\vec k_{\parallel}, z)\rangle$ are the BdG wavefunctions, and $E_n(\vec k_{\parallel}, z)$ are the corresponding eigenvalues with a band index $n$.\\ \newline

\begin{figure}[h!]
\includegraphics[width=1.7in,height=1.9in]{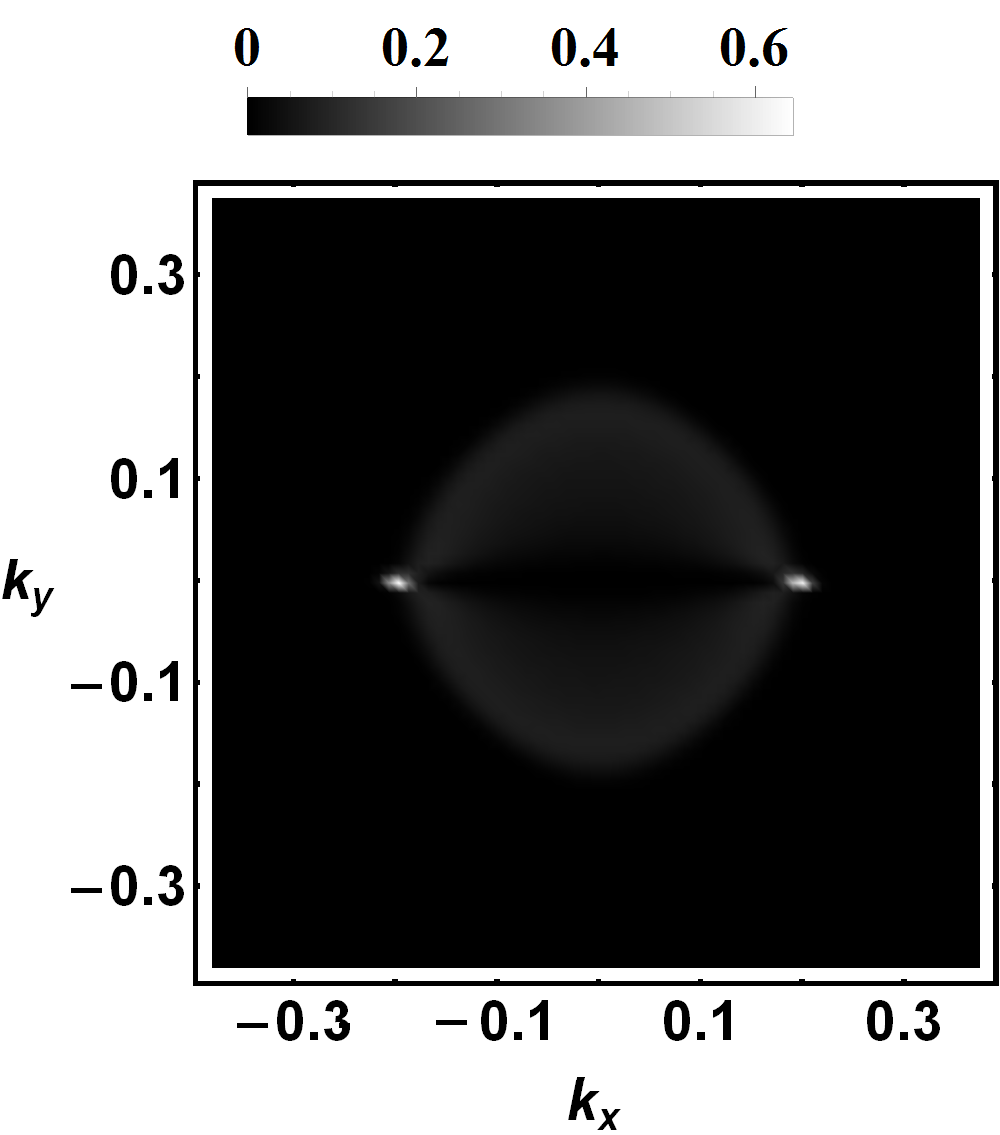}\hfill%
\includegraphics[width=1.7in,height=1.9in]{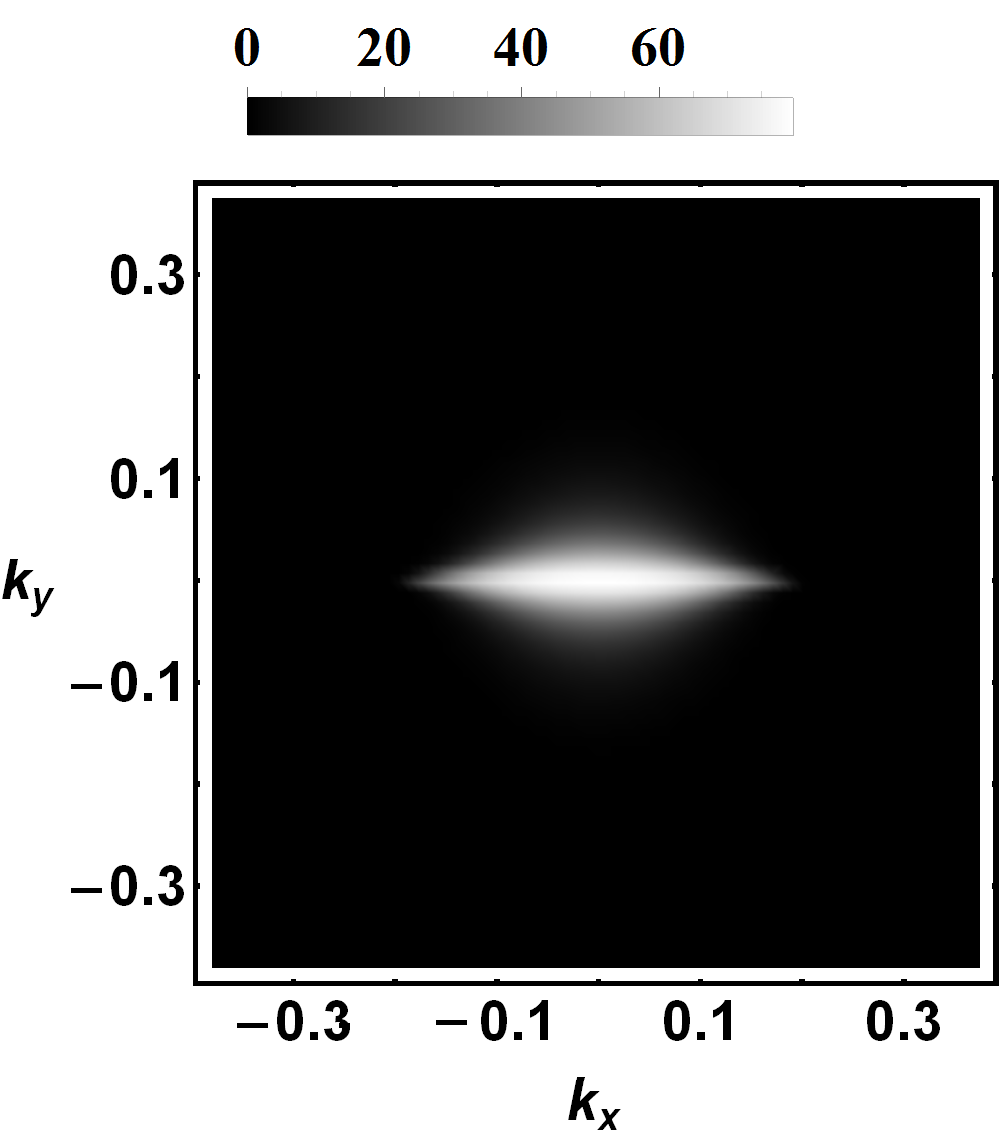}\hfill %
\includegraphics[width=1.7in,height=1.9in]{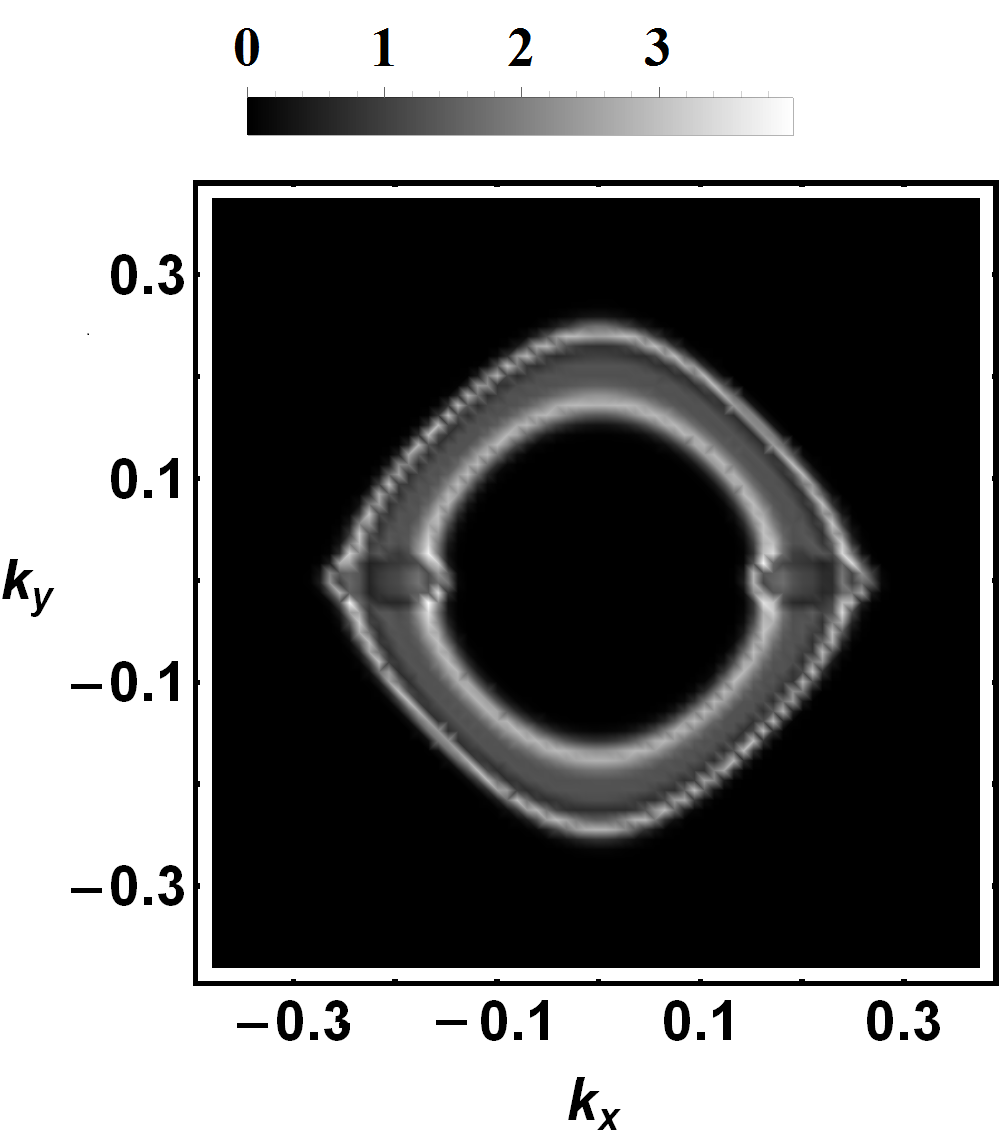}\hfill%
\includegraphics[width=1.7in,height=1.9in]{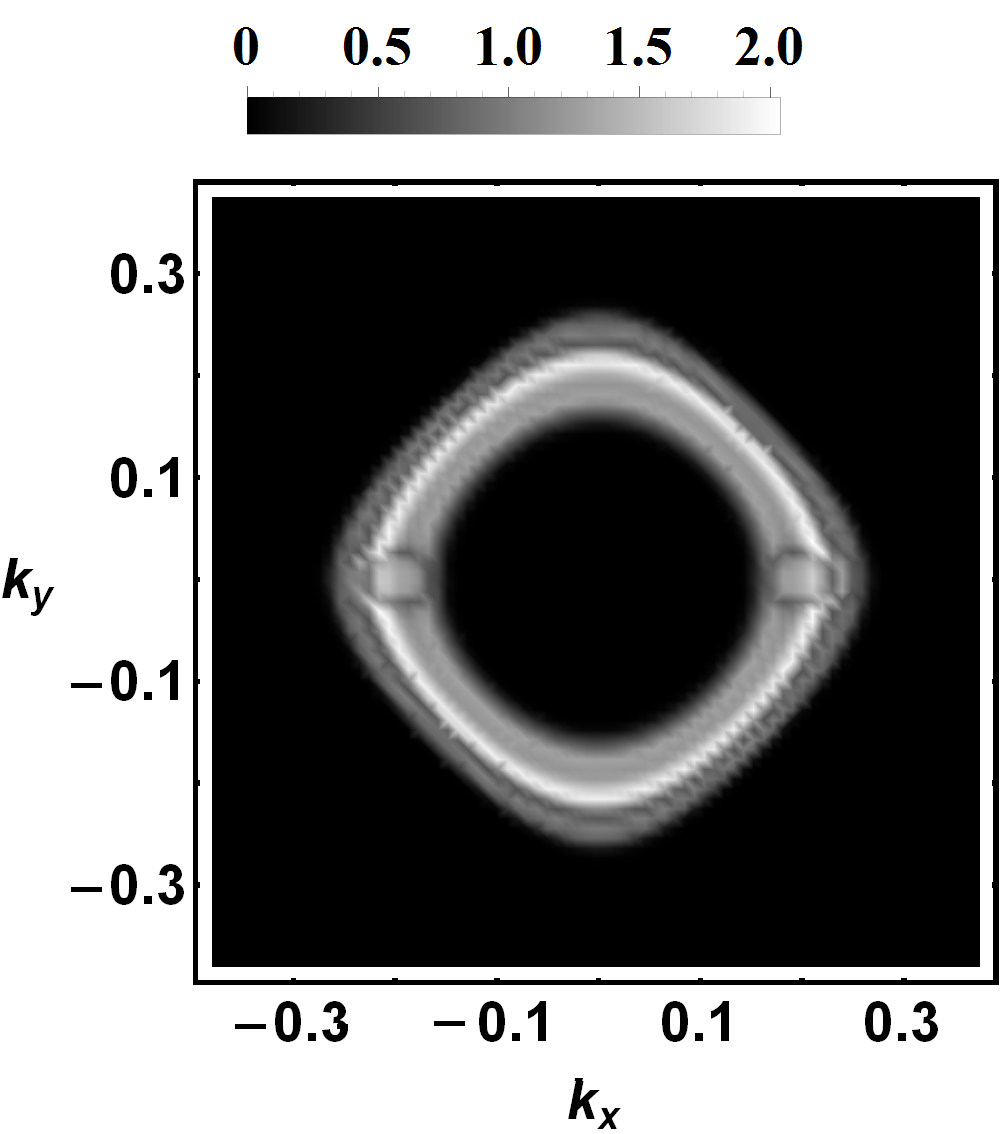}
\caption{Plots of the local density of states for a modified model (described in the main text) based on Ref.~\cite{Schnyder2016} but with a non-zero value of the perturbation parameter which converts the line nodal semimetal into a Weyl semimetal. (Left to right) Bulk density of states at zero energy, surface density of states at zero energy, bulk density of states at non-zero energy and surface density of states at non-zero energy. We have chosen the value of $\omega =0.1$ eV and the perturbation parameter to be $\delta = 0.1$ eV. }\label{EffectOfPerturbation}
\end{figure}

\textbf{\textit{Effect of a perturbation and non-zero frequency:} } In this section of the Supplemental Material, we study how it is possible to think of a point node semimetal as being the limiting case of a line nodal semimetal. It is possible to augment the Hamiltonian in Eq.~1 of the main text  to include the following term which smoothly interpolates between a line nodal semimetal and a Weyl semimetal
\begin{equation}
\hat{H}'(\vec k) =\delta \sin(a k_y) \tau_x \sigma_0,
\end{equation}
\noindent where $\delta$ controls the strength of the perturbation, $\sigma_i$ and $\tau_i$ are the Pauli matrices in the spin and orbital basis. Practically, it has been suggested that such a perturbation could be induced by light~\cite{yan2016tunable,narayan2016tunable,chan2016type,taguchi2016photovoltaic}. For the following discussion of the effect of such a perturbation term, refer to Fig. \ref{EffectOfPerturbation}.  At zero energy and a finite value of a perturbation parameter (chosen to be $\delta=0.1$ eV), the contour of Dirac points in the bulk of the line nodal semimetal shrinks into two Dirac points. On the surface, there exists a flat ``string'' of states (Fermi arcs), instead of a flat ``drum head'' like band, which connects the two point nodes. At finite energies, however, such a behavior is lacking due to the absence of the flat band, and one simply obtains a toroidal contour for the DOS in both the bulk and on the surface. \\
\begin{figure}[h!]
\includegraphics[width=1.7in,height=1.9in]{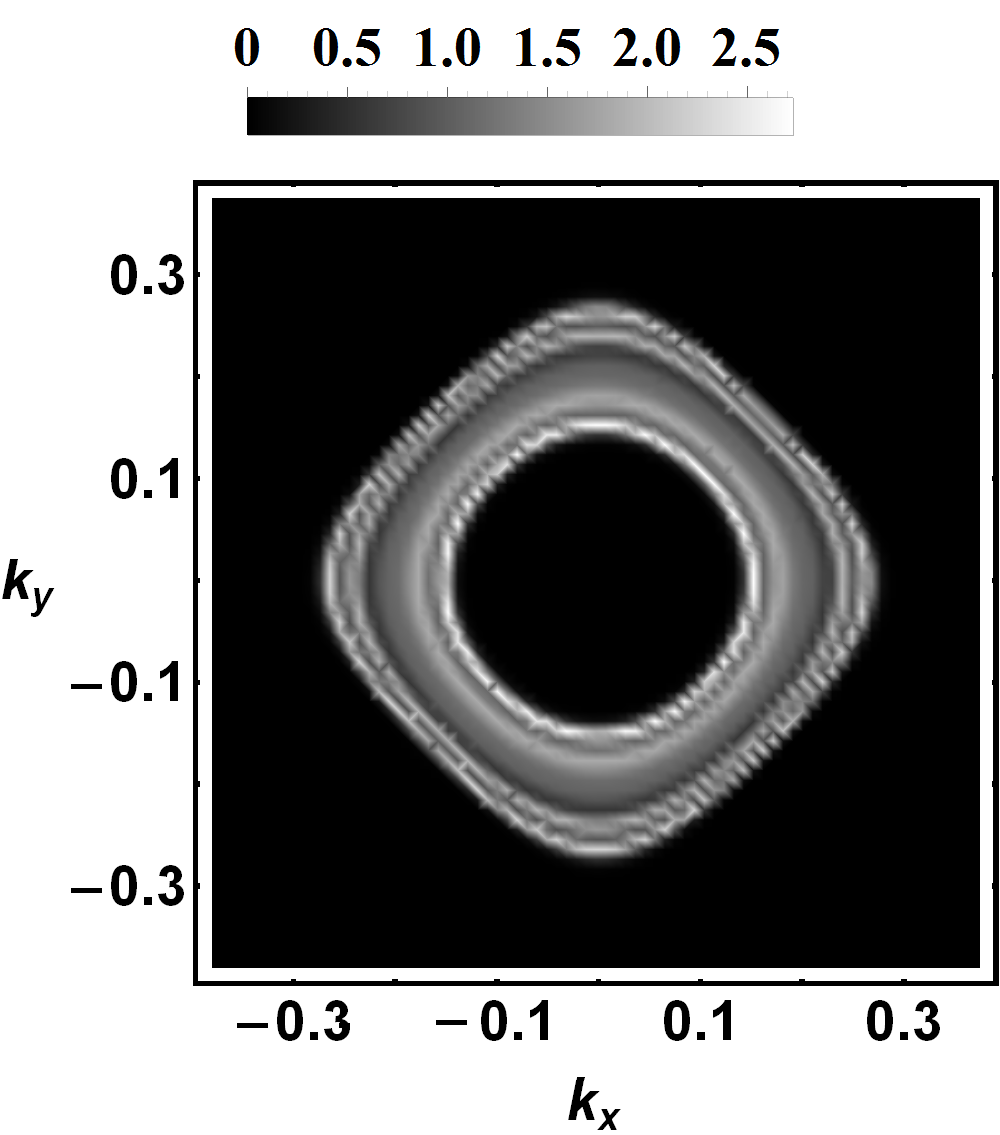}\hfill%
\includegraphics[width=1.7in,height=1.9in]{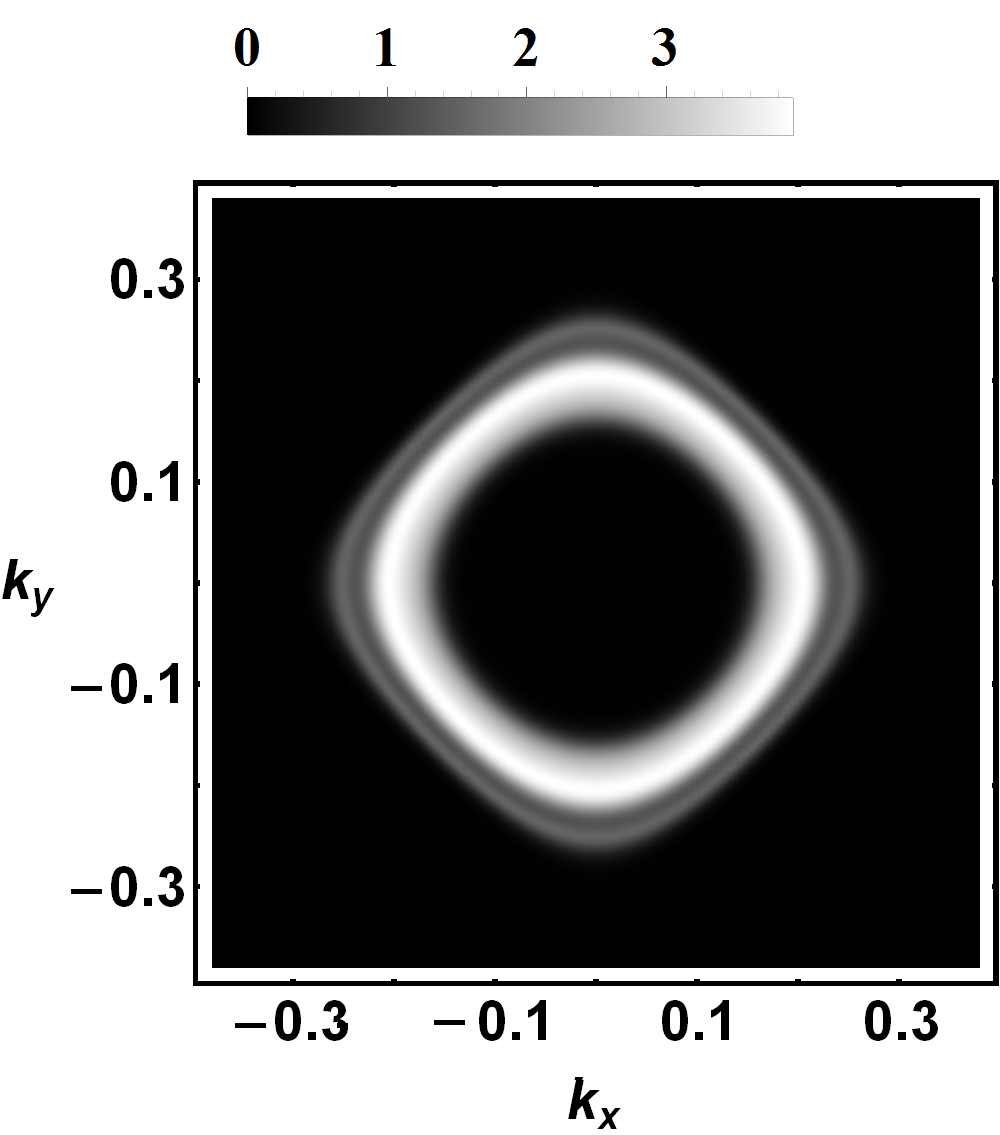}\hfill %
\includegraphics[width=1.7in,height=1.9in]{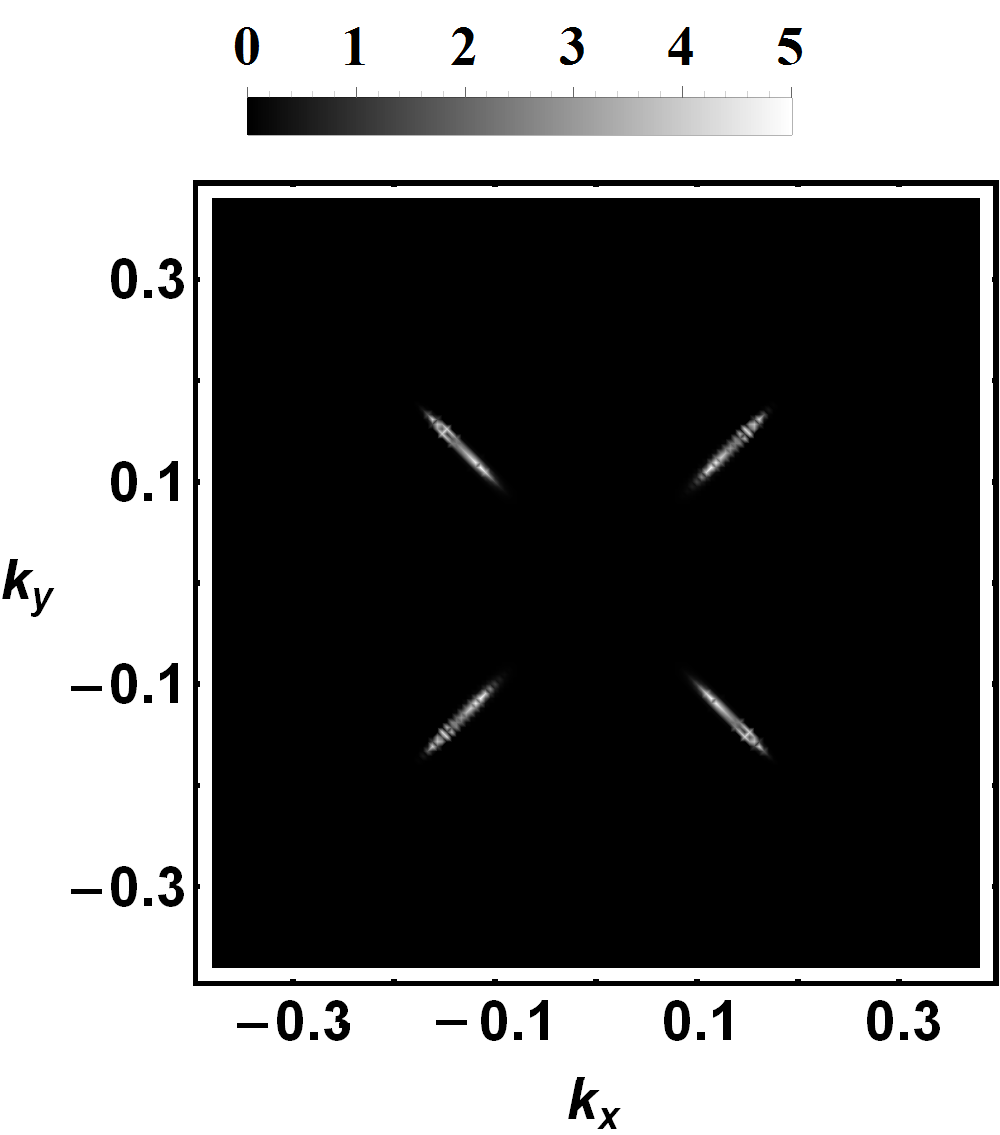}\hfill%
\includegraphics[width=1.7in,height=1.9in]{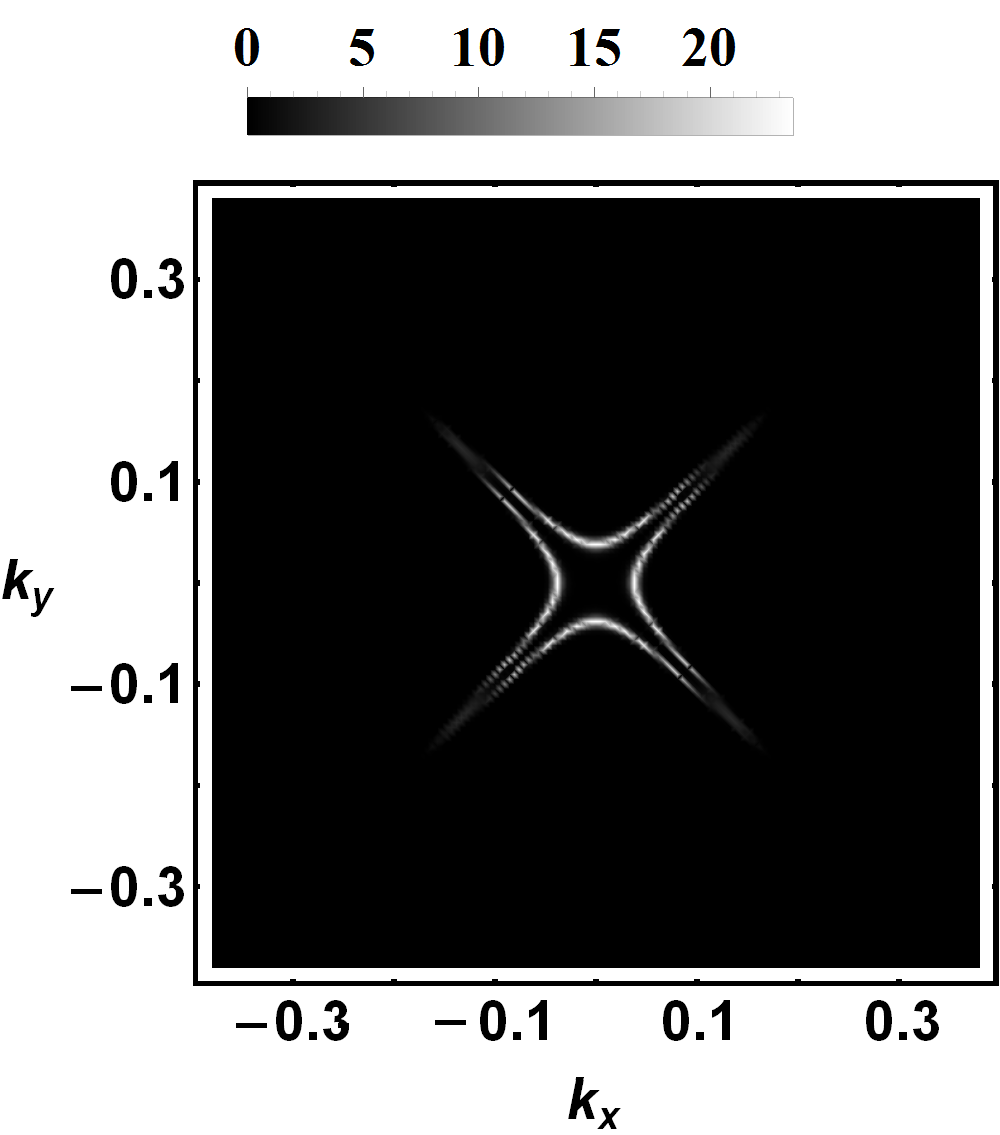}
\caption{Same as Fig.~1 appearing in the main text but at $\omega=0.1$ eV. Plots of the local density of states for the augmented model (described in the main text) based on Ref.~\cite{Schnyder2016}. (Left to right) Bulk density of states, surface density of states, bulk density of states in the superconducting state and surface density of states in the superconducting state. The perturbation parameter is set to $\delta = 0$. }\label{SchnyderModelNonZeroOmega}
\end{figure}
Fig.~\ref{SchnyderModelNonZeroOmega} shows the local density of states in the bulk and on the surface of a line nodal semimetal in the presence and absence of superconductivity at non-zero energies (similar to Fig.~1 in the main text but away from zero energy). In the bulk superconducting state, the four nodal points become slightly extended in momentum space along the diagonals of the Brillouin zone due to the toroidal band structure discussed in the main text. On the surface, however, when the induced superconducting gap is larger than frequency of the cut (chosen here to be $\omega = 0.1$ eV), i.e. when $\Delta_{0}>0.1$ eV, then all the surface bands are pushed above that frequency (at about the value of the superconducting gap) \textit{except} those states along the Brillouin zone diagonal. These states then converge down to the Fermi level to form line nodes at zero frequency, while at non zero frequency (less than the maximum value of superconducting gap) they form a ``petal" like structure shown in Fig.~2.\\ \newline

\textbf{\textit{Bound state calculations:}} We begin by recalling and expanding details of the $T$-matrix approximation that was used in the main text. The total Greens' function is written as
\begin{equation}
\hat G(\vec k, \vec k', \omega) =\hat G_0(\vec k, \omega) \delta_{\vec k, \vec k'} + \hat G_0(\vec k,\omega) \hat T(\vec k,\vec k', \omega) \hat G_0(\vec k', \omega).
\end{equation}
Here $G(\vec k, \vec k', \omega)$ and $G_0(\vec k, \omega)$  are the total interacting and non-interacting Greens functions and $T(\vec k, \vec k', \omega)$ is the $T$-matrix  which contains all the information about the impurity scattering. For the purposes of this article, we confine ourselves to scalar potential scatterers; this makes the $T$-matrix independent of momentum. Under this condition, the $T$-matrix becomes
\begin{eqnarray}
\hat T(\omega) &=& \hat V + \hat V \hat g_0(\omega)\hat V + \hat V \hat g_0(\omega)\hat V \hat g_0(\omega)\hat V +...\\ \nonumber
&=& \left[\hat{\sigma}_0 - \hat V \hat g_0(\omega)\right]^{-1}\hat V.
\end{eqnarray}
Here, we have defined 
\begin{equation}
\hat g_0(\omega) = \frac{1}{2\pi N_0}\sum_{\vec k} \hat G_0(\vec k, \omega),
\end{equation}
$N_0$ is the density of states at the Fermi level, and the scattering matrix $\hat V$ is given by $\frac{1}{c}\hat{\sigma}_3$. We have also used the parameter $c = \cot(N_0 U_0)$ as a measure of the strength of an $s$-wave scatterer as done in Ref.~\cite{Einzel1988}, where $U_0$ is the strength of the impurity scatterer; therefore, the unitarity limit (large scattering strength, $N_0 U_0\rightarrow \frac{\pi}{2}$) corresponds to the case when $c\rightarrow 0$.\\ \newline
\textit{Superconducting state with quadratic dispersion in two dimensions (D=2) and a $d-$wave pairing form:} For the sake of comparison with the case of the quasi-flat band, we revisit the calculation of Ref.~\cite{Einzel1988} for the energy  of the in gap bound state in a nodal, single band, $d-$wave superconductor. We begin by writing out its non-interacting Greens function given as 
\begin{equation}
\hat G_0(\vec k,\omega) = \left( \omega \hat{\sigma}_0 - \hat H_0(\vec k)\right)^{-1},
\end{equation}
where $\hat H_0(\vec k) = \epsilon(\vec k) \hat{\sigma}_3 + \Delta(\vec k)\hat{\sigma_1}$ and $\sigma_i$ are the Pauli matrices (henceforth, we will absorb the chemical potential $\mu$ into $\epsilon(\vec k)$ and keep it to be non-zero, in general. It will be explicitly shown where important). The Greens' function is explicitly evaluated as 
\begin{equation}
\hat G_0(\vec k, \omega) = \frac{1}{\Delta(\vec k)^2 + \epsilon(\vec k)^2 -\omega^2}
 \begin{pmatrix}
 -\left( \epsilon(\vec k) + \omega \right)&  -\Delta(\vec k)\\
  -\Delta(\vec k)& \left(\epsilon(\vec k) - \omega\right)
 \end{pmatrix}.
\end{equation}
We now proceed to evaluate the condition for the existence of a bound state when the Greens' function is a matrix. In general, the matrix $\hat g_0(\omega)$, can be written as
\begin{equation}
\hat g_0(\omega) = \frac{1}{2 \pi N_0} \sum_{\vec k} \hat G_0(\vec k, \omega) = \sum_{i}\mathscr G_i(\omega) \hat{\sigma}_i =
 \begin{pmatrix}
 \mathscr G_0(\omega) + \mathscr G_3(\omega)  & \mathscr G_1(\omega)   \\
\mathscr G_1(\omega)  &  \mathscr G_0(\omega)  - \mathscr G_3(\omega)
 \end{pmatrix},
\end{equation}
where the functions $\mathscr G_i(\omega)\equiv \frac{1}{2 \pi N_0}\sum_{\vec k} \mathscr G_i(\omega,\vec k)$, and $\mathscr{G}_0(\omega, \vec k) = \frac{-\omega}{D_{\vec k}}$, $\mathscr{G}_1(\omega, \vec k) = \frac{-\Delta(\vec k)}{D_{\vec k}}$, $\mathscr{G}_3(\omega, \vec k) = \frac{-\epsilon(\vec k)}{D_{\vec k}}$ and $D_{\vec k} = \Delta(\vec k)^2 + \epsilon(\vec k)^2 - \omega^2$. Given the form of the scattering matrix, $\hat V = \frac{1}{c}\hat{\sigma}_3$, the condition for the existence of bound states is that the determinant of $\left( \hat{\sigma}_0 - \hat V \hat g_0(\omega) \right)$ must vanish. As discussed in the main text, this condition is given by
\begin{equation}
\mathscr G_1(\omega)^2 - \mathscr G_0(\omega)^2 + (c - \mathscr G_3(\omega))^2 =0.
\end{equation}
Our task now is to evaluate these functions for the case of a $d-$wave superconductor with a quadratic dispersion in $D=2$. The function $\mathscr G_1(\omega)$ is zero since the gap function changes sign across the Brillouin zone and the $\phi$ integral vanishes. For $\mathscr G_3(\omega)$, we write
\begin{equation}
\mathscr G_3(\omega) = \frac{1}{2 \pi N_0} \sum_{\vec k}\frac{-\epsilon(\vec k)}{D_{\vec k}} = \frac{1}{2 \pi N_0} \left(\frac{L^2}{4 \pi^2}\right) \int_{-W}^{W} \frac{d\epsilon d\phi}{2 \alpha} \left[ \frac{-\epsilon}{\Delta_{\phi}^2 + \epsilon^2 - \omega^2}\right].
\end{equation}
Here, we have chosen a dispersion of the form $\epsilon(\vec k) = \alpha k^2 - \mu$ and $\Delta_{\phi} =  \Delta_0 \cos 2\phi$. From now on, we set the total bandwidth as $2W$ and a chemical potential ($\mu = E_f \sim  W$) close to or at half filling. As it can be seen, in the 2D case for a quadratic band, the chemical potential does not play a role. The integrand appearing above is anti-symmetric in $\epsilon$ and, hence, $\mathscr G_3(\omega) =0$. Next we calculate $\mathscr G_0(\omega)$ given by
\begin{equation}
\mathscr G_0(\omega) = \frac{1}{2 \pi N_0} \sum_{\vec k}\frac{-\epsilon(\vec k)}{D_{\vec k}} = \frac{1}{2 \pi N_0} \left(\frac{L^2}{4 \pi^2}\right) \int_{-W}^{W} \frac{d\epsilon d\phi}{2 \alpha} \left[ \frac{-\omega}{\Delta_{\phi}^2 + \epsilon^2 - \omega^2}\right],
\end{equation}
where $N_0$ is the total 2D density of states at the Fermi level and is given by $\frac{L^2}{4\pi \alpha}$. Performing the $\epsilon$ integral and substituting for $N_0$ yields
\begin{equation}
\mathscr G_0(\omega) = \frac{-1}{4 \pi} \int_0^{2\pi} \frac{d\phi}{\sqrt{\frac{\Delta_0^2 \cos^22\phi}{\omega^2} -1}},
\end{equation}
where we have substituted $\Delta_{\phi}$ for the $d$-wave order parameter and $\Delta_0$ is the pairing amplitude. The $\phi$ integral can be performed easily to give
\begin{equation}
\mathscr G_0(\omega) =\frac{-1}{4 \pi} \frac{4 \omega}{\sqrt{\Delta_0^2 -\omega^2}}  K\left(\frac{\Delta_0^2}{\Delta_0^2 -\omega^2}\right),
\end{equation}
where $K(x)$ is the elliptic $K$ function. Since we are looking for in gap bound states, we study the case where $\omega\ll \Delta_0$. A series expansion of the elliptic $K$ function is well known in this limit and $\mathscr G_0(\omega)$ reduces to 
\begin{equation}
\mathscr G_0(\omega) \simeq \frac{-\omega}{\pi \Delta_0}\left[ \log\left(\frac{4 \Delta_0}{\omega}\right) - i \frac{\pi}{2}\right].
\end{equation}
The condition for the existence of a bound state for this case simply reduces to $\mathscr G_0(\Omega) = \pm c$ as discussed in the main text. \\ \newline
\textit{Superconducting state with a quasi-flat band and a $d-$wave pairing form:} Here we aim to model the surface of a line-nodal semimetal and find its bound state properties. We choose a density of states profile as a Lorenztian of the form $\rho(\epsilon) = \frac{\gamma/\pi}{\gamma^2 + \epsilon^2}$, with a width $\gamma$, that peaks at the Fermi level. The energy scale $\gamma$ can be chosen to be the smallest among all other energy scales ($W,\Delta_0,\omega$) in the problem, as discussed in the main text. Just like the previous case, we have $\mathscr G_1(\omega)=0$ due to the $d-$wave sign change in the Brillouin zone. Moreover, the chosen density of states profile is even in $\epsilon$, $\mathscr G_3(\omega)$ is also zero. To calculate $\mathscr G_0(\omega)$, we substitute for the Lorentzian density of states profile into the momentum integral. Noting that the density of states at the fermi level diverges as $\gamma^{-1}$, we get 
\begin{equation}
\mathscr G_0(\omega) = \frac{\gamma^2}{2 \pi} \int_{-W}^{W} \frac{d\phi d\epsilon}{\gamma^2 + \epsilon^2}\left[ \frac{-\omega}{\Delta_{\phi}^2 + \epsilon^2 - \omega^2} \right].
\end{equation}
The $\epsilon$ integral can be performed to give
\begin{equation}
\mathscr G_0(\omega) = \frac{\gamma^2}{2 \pi} \int_{0}^{2\pi}\frac{2\omega d\phi}{\gamma}\left[ \frac{\arctan\left(\frac{W}{\gamma}\right) - \frac{\gamma}{\sqrt{\Delta_{\phi}^2 - \omega^2}} \arctan\left(\frac{W}{\sqrt{\Delta_{\phi}^2- \omega^2} }\right)}{\gamma^2 + \omega^2 - \Delta_{\phi}^2}\right].
\end{equation}
In the limit of large W (compared to the rest of the energy scales, $\gamma, \Delta_0,\omega$, with $\Delta_0>\omega$), the integral reduces to 
\begin{equation}
\mathscr G_0(\omega) \simeq \frac{\gamma^2}{2 \pi} \left(\frac{2\omega}{\gamma}\right) \int_{0}^{2\pi}\left[\frac{\frac{\pi}{2} - \frac{\pi \gamma}{2\sqrt{\Delta_{\phi}^2 - \omega^2}}}{-\Delta_{\phi}^2 + \gamma^2 + \omega^2}\right]d\phi.
\end{equation}
This integral can be performed and cast in terms of the function EllipticPi ($\Pi(x,y)$), i.e.
\begin{equation}
\mathscr G_0(\omega) \simeq \frac{\gamma^2}{2 \pi} \left(\frac{2\omega}{\gamma}\right)\left[\frac{-2\pi\gamma \Pi\left(\frac{\Delta_0^2}{\Delta_0^2 - \gamma^2 - \omega^2},\frac{\Delta_0^2}{\Delta_0^2 - \omega^2}\right)}{\sqrt{\Delta_0^2 - \omega^2}\left(\gamma^2 +\omega^2 -\Delta_0^2\right)}\right].
\end{equation}
We are interested in the limit where $\Delta_0\gg\omega\gg\gamma$. In this limit, $\mathscr G_0(\omega)$ reduces to 
\begin{equation}
\mathscr G_0(\omega) \simeq -\frac{2\gamma^2\xi}{\Delta_0^2} \left[\frac{1}{\xi^2} +\frac{1}{2}\log\left(\frac{\xi}{4}\right) + i\frac{\pi}{4}\right], 
\end{equation}
where we have defined $\xi\equiv\frac{\omega}{\Delta_0}$. This expression for $\mathscr{G}_0(\omega)$ has been used in the main text in obtaining Fig.~3. \\ \newline
\textit{Normal state with a quasi-flat band (Line nodal semi-metal surface):} We have not disscussed or summarized the normal state bound state properties in our manuscript as several works have already studied this in detail (See Ref.~\cite{Zhu2006} and references therein); however, we want to briefly state the result for the case of the quasi-flat band. To study the case of the flat band surface of a line nodal semi-metal without superconductivity we follow the same procedure as before. To this end, we consider the case of $\mu=0$ with the same Lorentzian density of states profile we used in the superconducting case. We obtain
\begin{equation}
\sum_{\vec k}G_0(\vec k, \omega) = \int_0^W \frac{\eta d\epsilon}{ \pi\left( \epsilon^2 + \eta^2 \right)\left(\omega - \epsilon\right)}.
\end{equation}
This integral can be performed without difficulty. In the limit of $\eta\rightarrow0$ we can obtain the bound state energy ($-\mid \omega \mid$) as 
\begin{equation}
\mid \omega \mid \simeq \frac{\mid U_0 \mid}{2}.
\end{equation}
Thus the bound state energy goes linearly with the strength of the impurity scatterer compared to quadratic for $D=1$ and exponential for $D=2$~\cite{Zhu2006}. \\ \newline
\textbf{\textit{Model for the line nodal semimetal:}} For the sake of completeness, we have provided the bulk band structure plots, bulk density of states, and the surface bands in the flat band and quasi-flat band limits in figures 3 and 4 of this Supplemental Material.\\ \newline
\begin{figure}[h!]
\includegraphics[width=3.6in,height=2.0in]{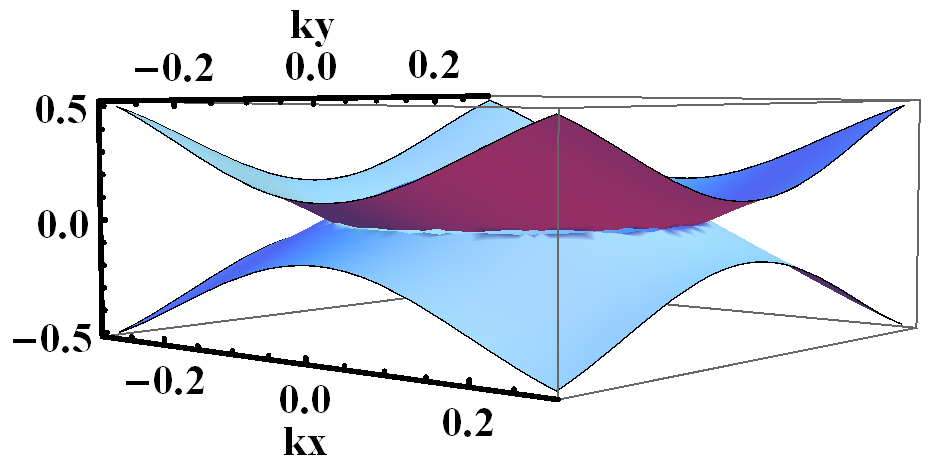}\hfill%
\includegraphics[width=3.0in,height=2.0in]{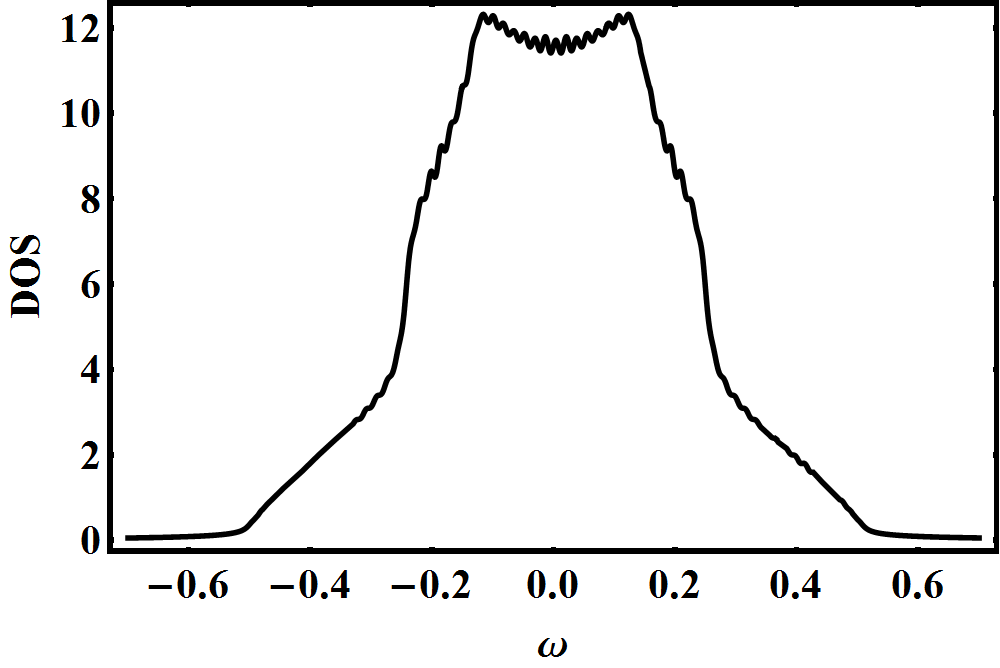}%
\caption{ Plots of the bulk line nodal bands (left) and bulk DOS (right) for the two band model appearing in the main text. }\label{BulkBandsAndDOS}
\end{figure}
\begin{figure}[h!]
\includegraphics[width=3.5in,height=2.5in]{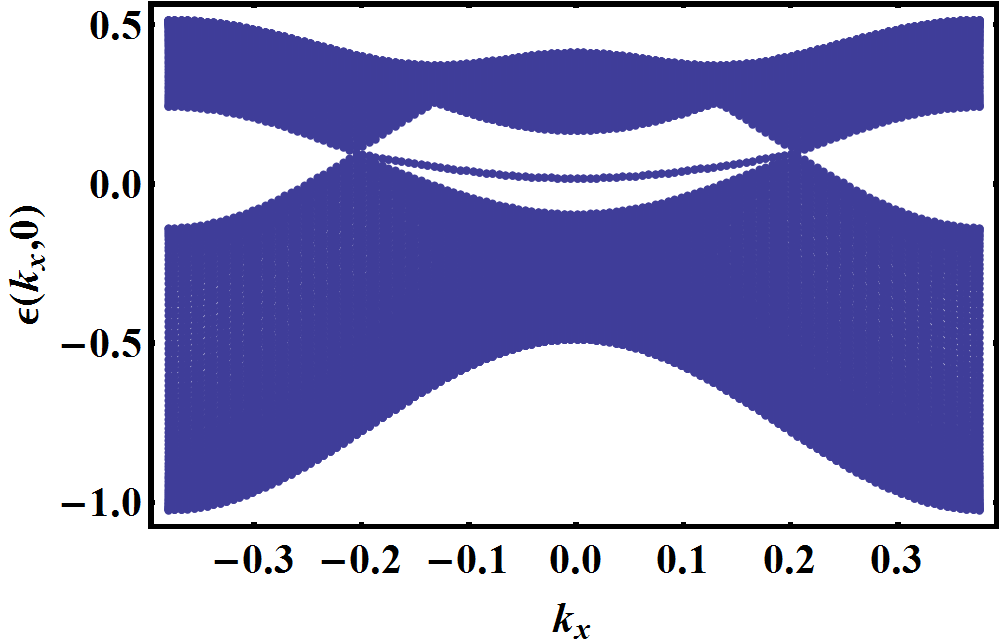}\hfill%
\includegraphics[width=3.5in,height=2.5in]{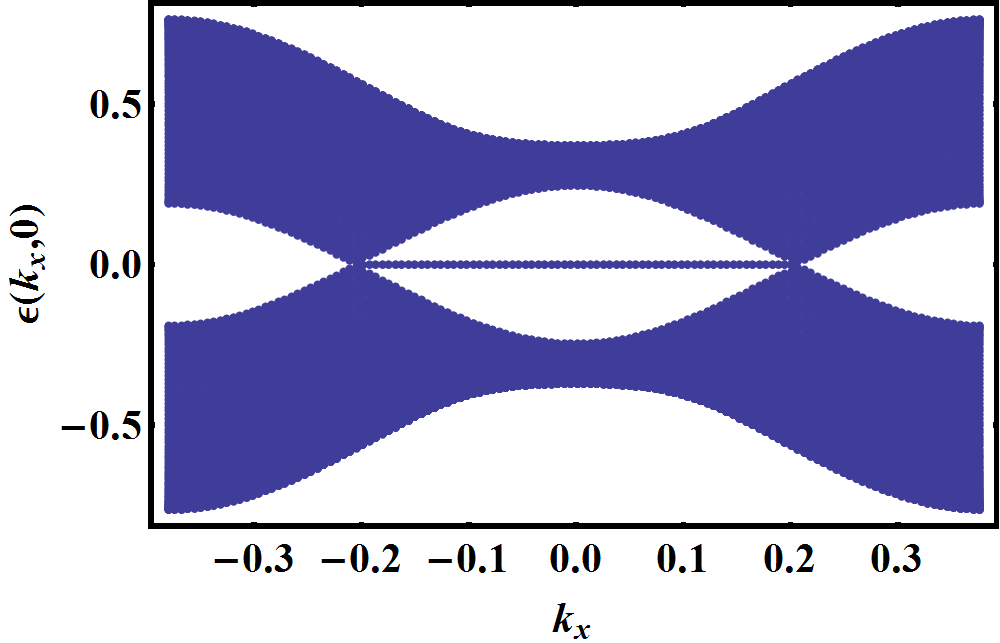}\hfill%
\caption{Energy dispersion along the $k_x$ axis with $k_y$ set to zero on the surface of the line nodal semimetal with open boundary conditions along the $z$ axis. (Left) Quasi-flat 'drum head' shaped surface band near the Fermi energy with $Z_0 = -0.156eV$.  (Right) Fully flat surface band with $Z_0 = 0.0$.  }\label{SurfaceBands}
\end{figure}
\end{document}